# Dense Error Correction via $\ell^1$-Minimization

John Wright, *Student Member,* and Yi Ma, *Senior Member.*

J. Wright and Y. Ma are with the Electrical and Computer Engineering Department, University of Illinois at Urbana-Champaign. Corresponding author: John Wright, 146 Coordinated Science Laboratory, 1308 West Main Street, Urbana, Illinois 61801, USA. Email: jnwright@uiuc.edu.

## Abstract

This paper studies the problem of recovering a non-negative sparse signal $\boldsymbol{x} \in \mathbb{R}^n$ from highly corrupted linear measurements $\boldsymbol{y} = A\boldsymbol{x} + \boldsymbol{e} \in \mathbb{R}^m$, where $\boldsymbol{e}$ is an unknown error vector whose nonzero entries may be unbounded. Motivated by an observation from face recognition in computer vision, this paper proves that for highly correlated (and possibly overcomplete) dictionaries $A$, any non-negative, sufficiently sparse signal $\boldsymbol{x}$ can be recovered by solving an $\ell^1$-minimization problem:

$$\min \|\boldsymbol{x}\|_1 + \|\boldsymbol{e}\|_1 \quad \text{subject to} \quad \boldsymbol{y} = A\boldsymbol{x} + \boldsymbol{e}.$$

More precisely, if the fraction $\rho$ of errors is bounded away from one and the support of $\boldsymbol{x}$ grows sublinearly in the dimension $m$ of the observation, then as $m$ goes to infinity, the above $\ell^1$-minimization succeeds for all signals $\boldsymbol{x}$ and almost all sign-and-support patterns of $\boldsymbol{e}$. This result suggests that accurate recovery of sparse signals is possible and computationally feasible even with nearly 100% of the observations corrupted. The proof relies on a careful characterization of the faces of a convex polytope spanned together by the standard crosspolytope and a set of iid Gaussian vectors with nonzero mean and small variance, which we call the "cross-and-bouquet" model. Simulations and experimental results corroborate the findings, and suggest extensions to the result.

## I. Introduction

Recovery of high-dimensional sparse signals or errors has been one of the fastest growing research areas in signal processing in the past few years. At least two factors have contributed to this explosive growth. On the theoretical side, the progress has been propelled by powerful tools and results from multiple mathematical areas such as measure concentration [1]–[3], statistics [4]–[6], combinatorics [7], and coding theory [8]. On the practical side, a lot of excitement has been generated by remarkable successes in real-world applications in areas such as signal (image or speech) processing [9], communications [10], computer vision and pattern recognition [11]–[13] etc.

### A. A Motivating Example

One notable, and somewhat surprising, successful application of sparse representation is automatic face recognition. As described in [11], face recognition can be cast as a sparse representation problem. For each person, a set of training images are taken under different illuminations. We can view each image as a vector by stacking its columns and put all the training images as column vectors of a matrix, say $A \in \mathbb{R}^{m\times n}$. Then, $m$ is the number of pixels in an image and $n$ is the total number of images for all the subjects of interest. Given a new query image, again we can stack it as a vector $\boldsymbol{y} \in \mathbb{R}^m$. To identify the image belongs to which subject, we can try to represent $\boldsymbol{y}$ as a linear combination of all the images, i.e., $\boldsymbol{y} = A\boldsymbol{x}$ for some $\boldsymbol{x} \in \mathbb{R}^n$. Since in practice $n$ can potentially be larger than $m$, the equations can be underdetermined and the solution $\boldsymbol{x}$ may not be unique. In this context, it is natural to seek the sparsest solution for $\boldsymbol{x}$ whose large non-zero coefficients then provide information about the subject's true identity. This can be done by solving the typical $\ell^1$-minimization problem:

$$\min_{\boldsymbol{x}} \|\boldsymbol{x}\|_1 \quad \text{subject to} \quad \boldsymbol{y} = A\boldsymbol{x}. \tag{1}$$

The problem becomes more interesting if the query image $\boldsymbol{y}$ is severely occluded or corrupted, as shown in Figure 1 left, column (a). In this case, one needs to solve a corrupted set of linear equations $\boldsymbol{y} = A\boldsymbol{x} + \boldsymbol{e}$, where $\boldsymbol{e} \in \mathbb{R}^m$ is an unknown vector whose nonzero entries correspond to the corrupted pixels. For sparse errors $\boldsymbol{e}$ and tall matrices $A$ ($m > n$), Candes and Tao [14] proposed to multiply the equation $\boldsymbol{y} = A\boldsymbol{x} + \boldsymbol{e}$ with a matrix $B$ such that $BA = 0$, and then use $\ell^1$-minimization to recover the error vector $\boldsymbol{e}$ from the new linear equation $B\boldsymbol{y} = B\boldsymbol{e}$.

As we mentioned earlier, in face recognition (and many other applications), $n$ can be larger than $m$ and the matrix $A$ can be full rank. One cannot directly apply the above technique even if the error $\boldsymbol{e}$ is known to be very sparse. To resolve this difficulty, in [11], the authors proposed to instead seek $[\boldsymbol{x}, \boldsymbol{e}]$ together as the sparsest solution to the extended equation $\boldsymbol{y} = [A\ \mathtt{I}]\boldsymbol{w}$ with $\boldsymbol{w} = \left[\begin{smallmatrix}\boldsymbol{x}\\ \boldsymbol{e}\end{smallmatrix}\right] \in \mathbb{R}^{m+n}$, by solving the extended $\ell^1$-minimization problem:

$$\min_{\boldsymbol{w}} \|\boldsymbol{w}\|_1 \quad \text{subject to} \quad \boldsymbol{y} = [A\ \ \mathtt{I}]\,\boldsymbol{w}. \tag{2}$$

This seemingly minor modification to the previous error correction approach has drastic consequences on the performance of robust face recognition. Solving the modified $\ell^1$-minimization enables almost perfect recognition even with more than 60% pixels of the query image are arbitrarily corrupted (see Figure 1 for an example), far beyond the amount of error that can theoretically be corrected by the previous error correction method [14].

Although $\ell^1$-minimization is expected to recover sufficiently sparse solutions with overwhelming probability for general systems of linear equations (see [16]), it is rather surprising that it works for the equation $\boldsymbol{y} = [A\ \mathtt{I}]\boldsymbol{w}$ at all. In the application described above, the columns of $A$ are highly correlated.

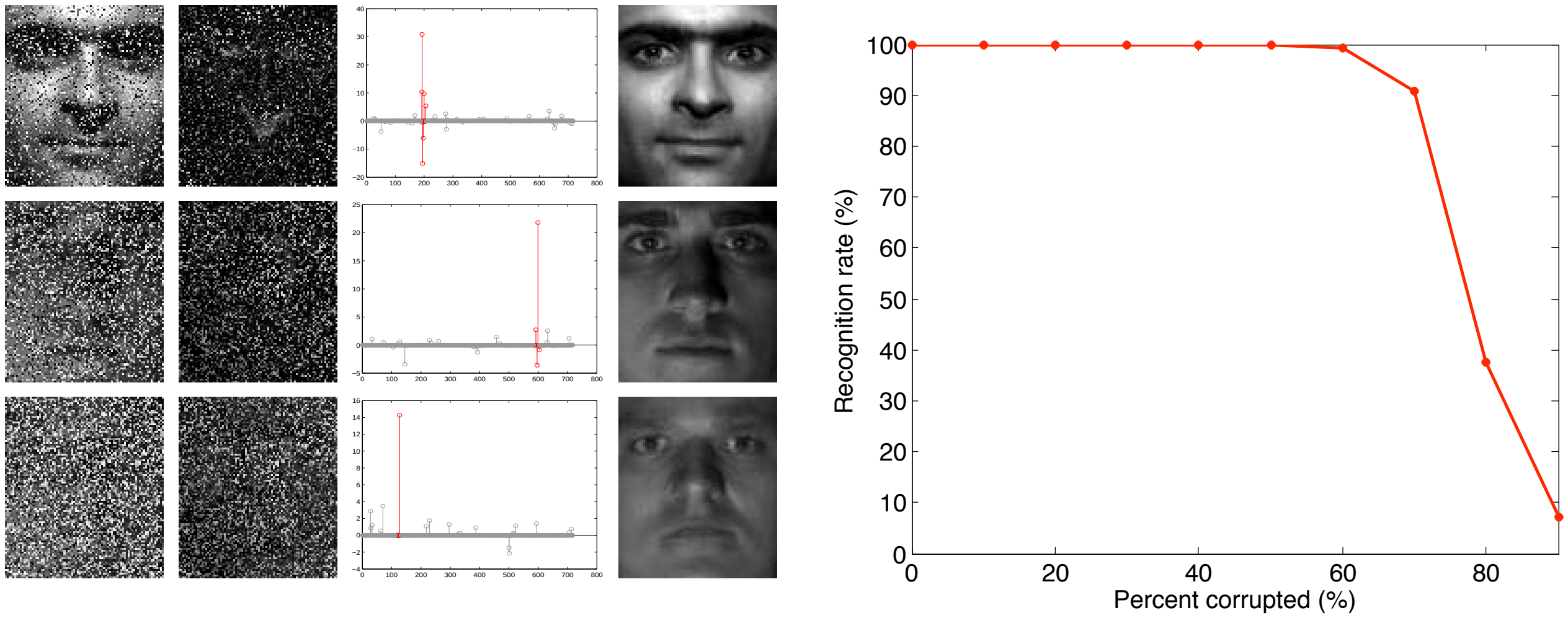

(a) (b) (c) (d)

Fig. 1

**Face recognition under random corruption.** LEFT: (A) TEST IMAGES $\boldsymbol{y}$ WITH RANDOM CORRUPTION FROM THE DATABASE PRESENTED IN [15]. TOP ROW: 30% OF PIXELS ARE CORRUPTED, MIDDLE ROW: 50% CORRUPTED, BOTTOM ROW: 70% CORRUPTED. (B) ESTIMATED ERRORS $\hat{\boldsymbol{e}}$. (C) ESTIMATED SPARSE COEFFICIENTS $\hat{\boldsymbol{x}}$. (D) RECONSTRUCTED IMAGES $\boldsymbol{y}_r = A\hat{\boldsymbol{x}}$. THE EXTENDED $\ell^1$-MINIMIZATION (2) CORRECTLY RECOVERS AND IDENTIFIES ALL THREE CORRUPTED FACE IMAGES. RIGHT: THE RECOGNITION RATE ACROSS THE ENTIRE RANGE OF CORRUPTION FOR ALL THE 38 SUBJECTS IN THE DATABASE. IT PERFORMS ALMOST PERFECTLY UPTO 60% RANDOM CORRUPTION.

As $m$ becomes large (i.e. the resolution of the image becomes high), the convex hull spanned by all face images of all subjects is only an extremely tiny portion of the unit sphere $\mathbb{S}^{m-1}$.[1] For example, the images in Figure 1 lie on $\mathbb{S}^{8,063}$. The smallest inner product with their normalized mean is 0.723; they are contained within a spherical cap of volume $\leq 1.47 \times 10^{-229}$. These vectors are tightly bundled together as a "bouquet," whereas the vectors associated with the identity matrix and its negative $\pm \mathtt{I}$ together[2] form a standard "cross" in $\mathbb{R}^m$, as illustrated in Figure 2. Notice that such a "cross-and-bouquet" matrix $[A\ \mathtt{I}]$ is neither incoherent nor (restrictedly) isometric, at least not uniformly. Also, the density of the desired solution $\boldsymbol{w}$ is not uniform either. The $\boldsymbol{x}$ part of $\boldsymbol{w}$ is usually a very sparse non-negative vector, but the $\boldsymbol{e}$ part can be very dense and have arbitrary signs. Existing results for recovering sparse signals

[1] At first sight, this seems somewhat surprising as faces of different people look so different to human eyes. That is probably because human brain has adapted to distinguish highly correlated visual signals such as faces or voices.

[2] Here we allow the entries of the error $\boldsymbol{e}$ to assume either positive or negative signs.

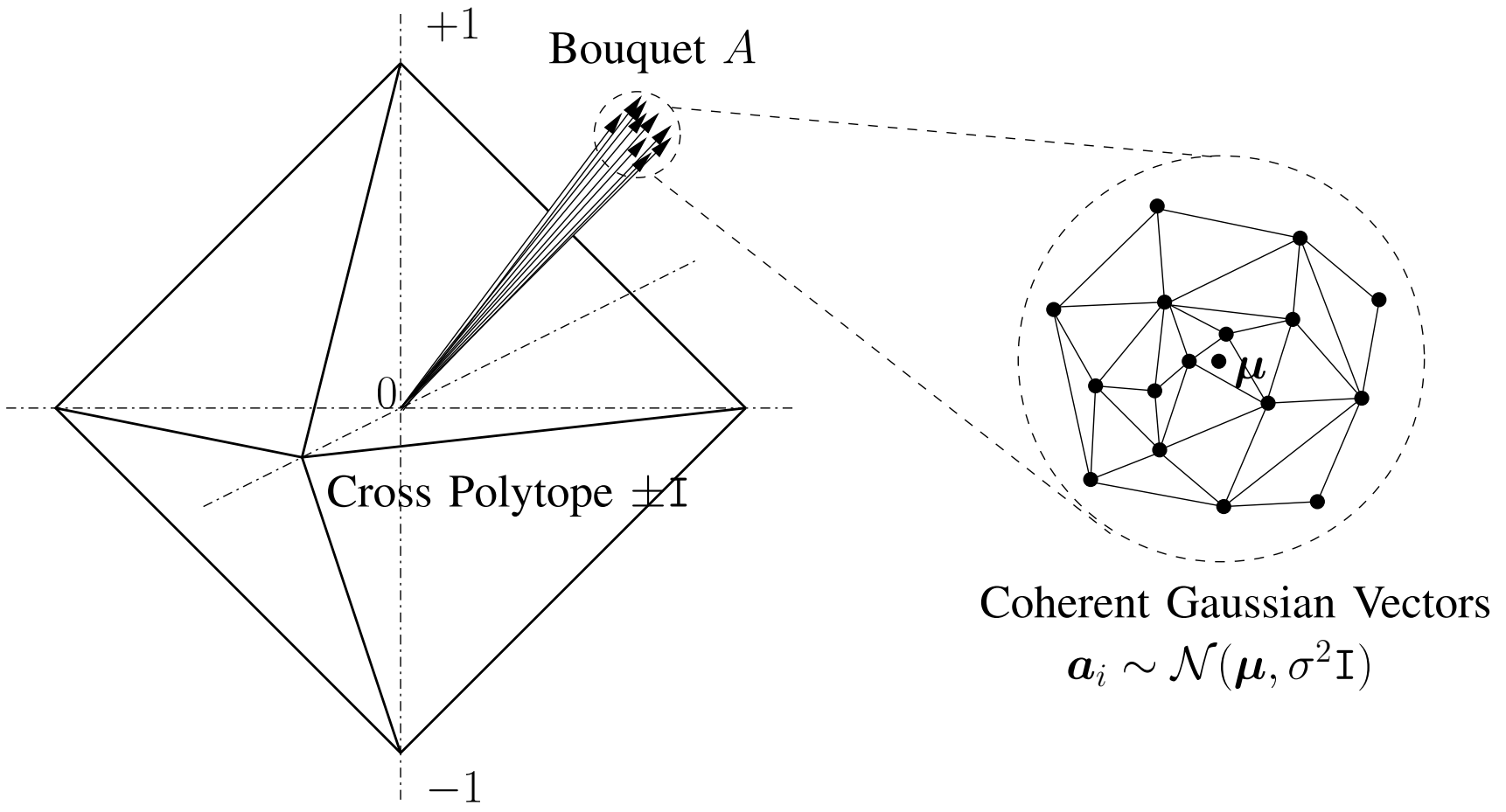

Fig. 2

**The "cross-and-bouquet" model.** LEFT: THE BOUQUET $A$ AND THE CROSSPOLYTOPE SPANNED BY THE MATRIX $\pm\mathtt{I}$. RIGHT: THE TIP OF THE BOUQUET MAGNIFIED; IT IS A COLLECTION OF IID GAUSSIAN VECTORS WITH SMALL VARIANCE $\sigma^2$ AND COMMON MEAN VECTOR $\boldsymbol{\mu}$. THE CROSS-AND-BOUQUET POLYTOPE IS SPANNED BY VERTICES FROM BOTH THE BOUQUET $A$ AND THE CROSS $\pm\mathtt{I}$.

suggest that $\ell^1$-minimization may have difficulty in dealing with such signals, contrary to its empirical success in face recognition.

We have experimented with similar cross-and-bouquet type models where the matrix $A$ is a random matrix with highly correlated column vectors. The simulation results in Section III indicate that what we have seen in face recognition is not an isolated phenomenon. In fact, the simulations reveal something even more striking and puzzling: As the dimension $m$ increases (and the sample size $n$ grows in proportion), the percentage of errors that the $\ell^1$-minimization (2) can correct seems to approach 100%! This may seem surprising, but this paper explains why this should be expected.

### *B. The Main Model and Result*

Motivated by the above empirical observations, this paper aims to resolve the apparent discrepancy between theory and practice of $\ell^1$-minimization and gives a more careful characterization of its behavior in recovering $[\boldsymbol{x}, \boldsymbol{e}]$ from the cross-and-bouquet (CAB) type models:

$$\boldsymbol{y} = A\boldsymbol{x} + \boldsymbol{e} = [\,A \ \ \mathtt{I}\,]\boldsymbol{w}. \tag{3}$$

We model the bouquet, the columns of $A$, as iid samples from a multivariate Gaussian distribution

$\mathcal{N}(\boldsymbol{\mu}, \sigma^2 \mathtt{I}_m)$, where $\sigma = \nu m^{-1/2}$ with $\nu$ sufficiently small, $\|\boldsymbol{\mu}\|_2 = 1$, and $\|\boldsymbol{\mu}\|_\infty \leq C_\mu m^{-1/2}$ for some $C_\mu \in \mathbb{R}_+$. These conditions insure that the bouquet remains tight as the dimension $m$ grows, and that its mean is mostly incoherent with the columns of the cross $\pm\mathtt{I}$.

We consider proportional growth for $m$ and $n$, that is, $n/m \to \delta \in \mathbb{R}_+$ as $m \to \infty$. However, the support size of the sparse signal $\boldsymbol{x}$ is only allowed to grow *sublinearly* in $m$: $\|\boldsymbol{x}\|_0 = O(m^{1-\eta})$ for some $\eta > 0$. This condition differs from (and is stronger than) the typical assumption in the sparse representation literature, where the support is often allowed to grow proportionally with the dimension [16]. In the next subsection, we will explain why the support of the signal $\boldsymbol{x}$ can only be sublinear if we allow the support of the error $\boldsymbol{e}$ to be arbitrarily dense. Nevertheless, this sublinear bound of sparsity is more than adequate for signals in many practical problems, including the face recognition problem. There, the support of $\boldsymbol{x}$ is bounded by a constant – the number of images per subject.

This paper proves that under the above conditions

> *for any $\rho < 1$, as $m$ goes to infinity, solving the $\ell^1$-minimization problem* (2) *correctly recovers any non-negative sparse signal $\boldsymbol{x}$ from almost any error $\boldsymbol{e}$ with support size $\leq \rho m$.*

We leave a more precise statement and the proof of the fact to Section II. In the remainder of this section, we discuss some of the main implications of this result in the broad context of sparse signal recovery, error correction, and some of its potential applications.

### C. Relations to Previous Results

*a) Restricted isometry and incoherence of the cross-and-bouquet model:* As mentioned earlier, typical results in the literature for sparse signal recovery do not apply to equations of the type $\boldsymbol{y} = A\boldsymbol{x} + \boldsymbol{e}$. The cross-and-bouquet matrix $[A\ \ \mathtt{I}]$ is neither highly isometric nor incoherent. As a result, greedy algorithms such as Orthogonal Matching Pursuit [17], [18] succeed only when the error $\boldsymbol{e}$ is very sparse (see Section III $a$) for the simulation results and comparison with our method). However, this does not mean that the restricted isometry property is irrelevant to the new problem. On the contrary, the proof of our results precisely rely on characterizing a special type of restricted isometry associated with this new problem, see Lemma 5 in Appendix A, which is used in the proof of our main result. Moreover, unlike the typical compressed sensing setting, the solution $[\boldsymbol{x}, \boldsymbol{e}]$ sought has very uneven density (or sparsity). This is reminiscent of the block sparsity studied in [19]. However, as we will see, the special block structure of the cross-and-boquet model enables sparse recovery far beyond the breakdown point for general sparse (or block sparse) signals.

*b) Error correction:* From an error correction viewpoint, the above result seems surprising: One can correctly solve a set of linear equations with *almost all* the equations randomly and arbitrarily corrupted! This is especially surprising considering that the best error-correcting codes (in the binary domain $\mathbb{Z}_2$), constructed based on expander graphs, normally correct a fixed fraction of errors [20]–[22]. The exact counterpart of our result in the binary domain is not clear.[3] While there are superficial similarities between our result and [21], [23] in the use of linear programming for decoding and analysis via polytope geometry, those works do not consider real valued signals. In particular, the negative result of [23] for specific families of binary codes admitting linear programming decoders does not apply here.

We can, however, draw the following comparisons with existing error correction methods in the domain of real numbers:

- When $n < m$, the range of $A$ is a subspace in $\mathbb{R}^m$. In such an overdetermined case, one could directly apply the method of Candes and Tao [14] mentioned earlier. However, the error vector $\boldsymbol{e}$ needs to be sparse for that approach whereas our result suggests even dense errors (with support far beyond 50%) can be corrected by instead solving the extended $\ell^1$-minimization (2). Thus, even in the overdetermined case, the new method has clear advantages for coherent matrices $A$. This will be verified by simulations in Section III $a$).
- The sublinear growth of the support of $\boldsymbol{x}$ in $m$ is the best one can hope for in the regime of dense errors. In general, we need at least $\|\boldsymbol{x}\|_0$ uncorrupted linear measurements to recover $\boldsymbol{x}$ uniquely. If an arbitrary fraction of the $m$ equations can be totally corrupted by $\boldsymbol{e}$, no fixed fraction of the equations remain good for recovering $\boldsymbol{x}$. If, on the other hand, the error $\boldsymbol{e}$ is sparse, then the $\ell^1$-minimization (2) is able to recover $\boldsymbol{x}$ with linear growth in support, as suggested by the existing theory [14], [16], [24]. Simulation results in Section III $d$) also confirm this phenomenon. However, in this paper, we are mainly interested in how the $\ell^1$-minimization behaves with dense errors, for $\rho \to 1$.
- When $n \geq m$, in general the Gaussian matrix $A$ is full rank and the method of Candes and Tao [14]

[3] It is possible that under an analogous growth model (see Section II-A), the LP decoder of [21] could also correct large fractions of binary errors.

no longer applies.[4] Our result suggests that as long as $A$ is highly correlated, the $\ell^1$-minimization (2) can still recover the sparse signal $\boldsymbol{x}$ correctly even if almost all the equations might be corrupted. This is verified by the simulation results in Section III $c$).

*c) Polytope geometry:* The success of $\ell^1$-minimization in recovering sparse solutions $\boldsymbol{x}$ from under-determined systems of linear equations $\boldsymbol{y} = A\boldsymbol{x}$ can be viewed as a consequence of a surprising property of high-dimensional polytopes. If the column vectors of $A$ are random samples from a zero-mean Gaussian $\mathcal{N}(0, \mathtt{I})$, and $m$ and $n$ are allowed to grow proportionally, then with overwhelming probability the convex polytope conv$(A)$ spanned by the columns of $A$ is highly neighborly [24], [25]. Neighborliness provides the necessary and sufficient condition for uniform sparse recovery: the $\ell^1$-minimization (1) correctly recovers $\boldsymbol{x}$ if and only if the columns associated with the nonzero entries of $\boldsymbol{x}$ span a face of the polytope conv$(A)$.

In our case, the columns of the matrix $A$ are iid Gaussian vectors with nonzero mean $\boldsymbol{\mu}$ and small variance $\sigma^2$, whereas the vectors of the cross $\pm\mathtt{I}$ are completely fixed. To characterize when the extended $\ell^1$-minimization (2) is able to recover the solution $[\boldsymbol{x}, \boldsymbol{e}]$ correctly, we need to examine the geometry of the peculiar convex polytope conv$(A, \pm\mathtt{I})$ spanned together by the random bouquet $A$ and the fixed cross $\pm\mathtt{I}$. Thus, it comes as no surprise that the proof of our main result relies on a careful study of the geometry of such a "cross-and-bouquet" polytope. As we will show that indeed, the vertices associated with the non-zero entries of $\boldsymbol{x}$ and $\boldsymbol{e}$ form a face of the polytope with probability approaching one as the dimension $m$ becomes large. Precisely due to high neighborliness of the cross-and-bouquet polytopes, the extended $\ell^1$-minimization (2) is able to correctly recover the desired solution, even though the part of the solution corresponding to $\boldsymbol{e}$ might be dense.

### D. Implications on Applications

*a) Robust reconstruction, classification, and source separation:* The new result about the cross-and-bouquet model has strong implications on robust reconstruction, classification, and separation of highly correlated classes of signals such as faces or voices, despite severe corruption. It helps explain the surprising performance of face recognition that we discussed earlier. It further suggests that if the resolution of the image increases in proportion with the size of the database (say, due to the increasing number of subjects), the $\ell^1$-minimization would tolerate even higher level of corruption, far beyond the 60% at the resolution experimented with in [11]. Other applications where this kind of model could be useful and effective include speech recognition/imputation, audio source separation, video segmentation, or activity recognition from motion sensors.

[4]One could choose to pre-multiply the equation $\boldsymbol{y} = A\boldsymbol{x} + \boldsymbol{e}$ with an "approximate orthogonal complement" of $A$, say the orthogonal complement of the mean vector $\boldsymbol{\mu}$, which is an $(m-1) \times m$ matrix $B$. Then the equation becomes $B\boldsymbol{y} = B\boldsymbol{e} + \boldsymbol{z}$ where $\boldsymbol{z} = BA\boldsymbol{x}$. If the norm of $\boldsymbol{x}$ is bounded, then $\boldsymbol{z}$ is a signal with small magnitude due to the near-orthogonality of $B$ and $A$. In this case, one can view $\boldsymbol{z}$ as a noise term and try to recover $\boldsymbol{e}$ as a sparse signal via $\ell^1$-minimization. However, for $\boldsymbol{e}$ with arbitrary signs, the breakdown point for such $\ell^1$-minimization is less than $50\%$.

*b) Communication through an almost random channel:* The result suggests that we can use the cross-and-bouquet model to accurately send information through a highly corrupting channel. Hypothetically, we can imagine a channel through which we can send one real number at a time, say as one packet of binary bits, and each packet has a high probability of being totally corrupted. One can use the sparse vector $\boldsymbol{x}$ (or its support) to represent useful information, and use a set of highly correlated high-dimensional vectors as the encoding transformation $A$. The high correlation in $A$ ensures that there is sufficient redundancy built in the encoded message $A\boldsymbol{x}$ so that the information about $\boldsymbol{x}$ will not be lost even if many entries of $A\boldsymbol{x}$ can be corrupted while being sent through such a channel. Our result suggests that the decoding can be done correctly and efficiently using linear programming.

*c) Encryption and information hiding:* One can potentially use the cross-and-bouquet model for encryption. For instance, if both the sender and receiver share the same encoding matrix $A$ (say a randomly chosen Gaussian matrix), the sender can deliberately corrupt the message $A\boldsymbol{x}$ with arbitrary random errors $\boldsymbol{e}$ before sending it to the receiver. The receiver can use linear programming to decode the information $\boldsymbol{x}$, whereas any eavesdropper will not be able to make much sense out of the highly corrupted message $\boldsymbol{y} = A\boldsymbol{x} + \boldsymbol{e}$. Of course, the long-term security of such an encryption scheme relies on the difficulty of learning the encoding matrix $A$ after gathering many instances of corrupted message. It is not even clear whether it is easy to learn $A$ from instances of uncorrupted message $\boldsymbol{y} = A\boldsymbol{x}$. Even if the dimensions of the matrix $A$ are given, effectively learning $A$ from a set of observed messages $\boldsymbol{Y} = [\boldsymbol{y}_1, \boldsymbol{y}_2, \ldots, \boldsymbol{y}_k]$ is still a largely open problem, known in the literature as the "dictionary learning" problem. Existing algorithms are iterative or greedy in nature, with no guarantee of global optimality [9]. Although its hardness has not been precisely characterized, we expect dictionary learning from highly corrupted observations to be an even more daunting problem, a challenge for anyone who tries to break this encryption scheme.

## II. ROADMAP OF THE PROOF

In this section, we begin with a precise statement of our main result in Section II-A. We then lay out the roadmap for the proof. Section II-B outlines the key geometric picture behind the proof. In Section II-C, we then prove the main result, assuming that two technical conditions in Lemma 2 hold. Section II-D discusses the ideas required to establish these conditions, leaving a number of details to the Appendix.

### A. Problem Statement

Motivated by the face recognition example introduced above, we consider the problem of recovering a non-negative[5] sparse signal $\boldsymbol{x}_0 \in \mathbb{R}^n$ from highly corrupted observations $\boldsymbol{y} \in \mathbb{R}^m$:

$$\boldsymbol{y} = A\boldsymbol{x}_0 + \boldsymbol{e}_0,$$

where $\boldsymbol{e}_0 \in \mathbb{R}^m$ is a sparse vector of errors of arbitrary magnitude. The model for $A \in \mathbb{R}^{m\times n}$ should capture the idea that it consists of small deviations about a mean, hence a "bouquet." In this paper, we consider the case where the columns of $A$ are iid samples from a Gaussian distribution:

$$A = [\boldsymbol{a}_1, \dots, \boldsymbol{a}_n] \in \mathbb{R}^{m\times n}, \quad \boldsymbol{a}_i \sim_{iid} \mathcal{N}\left(\boldsymbol{\mu}, \frac{\nu^2}{m}\mathrm{I}_m\right), \quad \|\boldsymbol{\mu}\|_2 = 1, \quad \|\boldsymbol{\mu}\|_\infty \le C_\mu m^{-1/2}. \tag{4}$$

Together, the two assumptions on the mean force it to remain incoherent with the standard basis (or "cross") as the dimension increases.

We study the behavior of the solution to the $\ell^1$-minimization (2) in this model, in the following asymptotic framework, which we term "weak proportional growth":

*Assumption 1 (Weak Proportional Growth):* A sequence of signal-error problems exhibits weak proportional growth with parameters $\delta > 0, \rho \in (0,1), C_0 > 0, \eta_0 > 0$, denoted $\mathrm{WPG}_{\delta,\rho,C_0,\eta_0}$ if as $m \to \infty$,

$$\frac{n}{m} \to \delta, \qquad \frac{\|\boldsymbol{e}_0\|_0}{m} \to \rho, \qquad \|\boldsymbol{x}_0\|_0 \le C_0\, m^{1-\eta_0}. \tag{5}$$

This should be contrasted with the "total proportional growth" (TPG) setting of, e.g., [26], in which the number of nonzero entries in the signal $\boldsymbol{x}_0$ also grows as a fixed fraction of the dimension. In that setting, one might expect a sharp phase transition in the combined sparsity of $(\boldsymbol{x}_0, \boldsymbol{e}_0)$ that can be recovered by $\ell^1$-minimization.[6] In WPG, on the other hand, we observe a striking phenomenon not seen in TPG: the correction of arbitrary fractions of errors. This comes at the expense of the stronger assumption that $\|\boldsymbol{x}_0\|_0 = o(m)$, an assumption that is valid in some real applications such as the face recognition example above.

[5]The non-negativity assumption is important: in the highly-coherent systems considered here, $\ell^1$-minimization generally does not recover signals $\boldsymbol{x}_0$ with arbitrary signs. Geometrically, this is would require vectors from the "bouquet" to "see" through the crosspolytope to vectors that are nearly antipodal to them.

[6]Existing results (e.g., [24]) do not prove the existence of phase transitions in inhomogeneous models such as the one considered here. However, simulations suggest that in total proportional growth, such transitions do occur (see Section III $d$)).

Before stating our main result, we fix some additional notation. For any $n \in \mathbb{Z}_+$, $[n]$ denotes the set $\{1, \dots, n\}$. Let $I = \mathrm{supp}(\boldsymbol{x}_0) \subset [n]$, $J = \mathrm{supp}(\boldsymbol{e}_0) \subset [m]$, $\boldsymbol{\sigma} = \mathrm{sgn}(\boldsymbol{e}_0(J))$, and let $k_1 = |I|$ be the support size of the signal $\boldsymbol{x}_0$ and $k_2 = |J|$ the support size of the error $\boldsymbol{e}_0$. For an arbitrary $r_1 \times r_2$ matrix $M$, if $L_1 \subset [r_1]$ and $L_2 \subset [r_2]$, $M_{L_1,L_2}$ denotes the $|L_1| \times |L_2|$ submatrix of $A$ indexed by these quantities. We use $M_{L_1,\bullet}$ as a shorthand for $M_{L_1,[r_2]}$. $M^*$ denotes the transpose of $M$. Also, we use $\mathbf{1}_I$ (or $\mathbf{1}_J$) to represent a vector in $\mathbb{R}^n$ (or $\mathbb{R}^m$) that has ones on the support $I$ (or $J$) and zeros elsewhere. To reduce confusion between the index set $I$ and the identity matrix, we use $\mathtt{I}$ to denote the latter. Below, where the symbol $C$ occurs with no subscript, it should be read as "some constant." When used in different sections, it need not refer to the same constant.

In the following, we say the cross-and-bouquet model is *$\ell^1$-recoverable at $(I, J, \boldsymbol{\sigma})$* if for all $\boldsymbol{x}_0 \geq 0$ with support $I$ and $\boldsymbol{e}_0$ with support $J$ and signs $\boldsymbol{\sigma}$, we have

$$(\boldsymbol{x}_0, \boldsymbol{e}_0) = \arg\min \|\boldsymbol{x}\|_1 + \|\boldsymbol{e}\|_1 \quad \text{subject to} \quad A\boldsymbol{x} + \boldsymbol{e} = A\boldsymbol{x}_0 + \boldsymbol{e}_0, \tag{6}$$

and the minimizer is uniquely defined. From the geometry of $\ell^1$-minimization, if (6) does not hold for some pair $(\boldsymbol{x}_0, \boldsymbol{e}_0)$, then it does not hold for any $(\boldsymbol{x}, \boldsymbol{e})$ with the same signs and support as $(\boldsymbol{x}_0, \boldsymbol{e}_0)$ [25]. Understanding $\ell^1$-recoverability at each $(I, J, \boldsymbol{\sigma})$ completely characterizes which solutions to $\boldsymbol{y} = A\boldsymbol{x} + \boldsymbol{e}$ can be correctly recovered. In this language, our main result can be stated more precisely as:

*Theorem 1 (Error Correction with the Cross-and-Bouquet Model):* For any $\delta > 0$, $\exists\, \nu_0(\delta) > 0$ such that if $\nu < \nu_0$ and $\rho < 1$, in $\mathrm{WPG}_{\delta,\rho,C_0,\eta_0}$ with $A$ distributed according to (4), error support $J$ chosen uniformly at random from $\binom{[m]}{k_2}$ and error signs $\boldsymbol{\sigma}$ chosen uniformly at random from $\{\pm 1\}^{k_2}$,

$$\lim_{m \to \infty} P_{A,J,\sigma}\left[\, \ell^1\text{-recoverability at } (I, J, \boldsymbol{\sigma}) \;\; \forall\, I \in \binom{[n]}{k_1} \,\right] \;=\; 1. \tag{7}$$

In other words, as long as the bouquet is sufficiently tight, asymptotically $\ell^1$-minimization recovers any non-negative sparse signal from almost any error with support size less than 100%.

## B. Problem Geometry

We first restate the necessary and sufficient conditions for $\ell^1$-recoverability geometrically, as separation of a higher-dimensional $\ell^1$-ball and an affine subspace (see Figure 3). To witness this separation, we must show the existence of a separating hyperplane, whose normal we will denote by $\boldsymbol{q}$.

*Lemma 1:* Fix $(I, J, \boldsymbol{\sigma})$, and define $\boldsymbol{w} \doteq A^*_{J,\bullet}\boldsymbol{\sigma} - \mathbf{1}_I \in \mathbb{R}^n$ and

$$G \doteq \begin{bmatrix} A_{J^c,I} & A_{J^c,I^c} \\ 0 & \mathtt{I}_{n-k_1} \end{bmatrix} \in \mathbb{R}^{p \times n}, \quad p = m + n - k_1 - k_2. \tag{8}$$

Suppose $G$ has full column rank $n$.[7] The model is $\ell^1$-recoverable at $(I, J, \boldsymbol{\sigma})$ iff

$$\exists\, \boldsymbol{q} \in \mathbb{R}^p \quad \text{such that} \quad \|\boldsymbol{q}\|_\infty < 1 \quad \text{and} \quad G^* \boldsymbol{q} = \boldsymbol{w}. \tag{9}$$

*Proof:* As above, let $\boldsymbol{y} = A\boldsymbol{x}_0 + \boldsymbol{e}_0$. The pair $(\boldsymbol{x}_0, \boldsymbol{e}_0)$ is the unique minimum $\ell^1$-norm solution to the equation $\boldsymbol{y} = A\boldsymbol{x} + \boldsymbol{e}$ iff

$$\nexists\, (\Delta\boldsymbol{x}, \Delta\boldsymbol{e}) \neq \mathbf{0} \,:\, A\Delta\boldsymbol{x} = -\Delta\boldsymbol{e},\ \|\boldsymbol{x} + \Delta\boldsymbol{x}\|_1 + \|\boldsymbol{e} + \Delta\boldsymbol{e}\|_1 \leq \|\boldsymbol{x}\|_1 + \|\boldsymbol{e}\|_1. \tag{10}$$

Due to the geometry of $\ell^1$-minimization and the convexity of $\|\cdot\|_1$, we lose no generality in assuming that $\boldsymbol{x} = \mathbf{1}_I$, $\boldsymbol{e} \in \{-1, 0, 1\}^m$ and $\|\Delta\boldsymbol{x}\|_\infty < 1$, $\|\Delta\boldsymbol{e}\|_\infty < 1$. Then,

$$\|\boldsymbol{x} + \Delta\boldsymbol{x}\|_1 = \|\boldsymbol{x}\|_1 + \mathbf{1}_I^* \Delta\boldsymbol{x} + \|\Delta\boldsymbol{x}_{I^c}\|_1, \quad \text{and} \quad \|\boldsymbol{e} + \Delta\boldsymbol{e}\|_1 = \|\boldsymbol{e}\|_1 + \boldsymbol{e}^* \Delta\boldsymbol{e} + \|\Delta\boldsymbol{e}_{J^c}\|_1.$$

Substituting into (10) and using $\Delta\boldsymbol{e} = -A\Delta\boldsymbol{x}$ yields that $(\boldsymbol{x}, \boldsymbol{e})$ is optimal iff

$$\nexists\, \Delta\boldsymbol{x} \neq \mathbf{0} \;:\; \|A_{J^c,\bullet}\Delta\boldsymbol{x}\|_1 + \|\Delta\boldsymbol{x}_{I^c}\|_1 \leq \langle A^*\boldsymbol{e} - \mathbf{1}_I, \Delta\boldsymbol{x}\rangle.$$

Condition (II-B) is satisfied iff

$$\forall\, \Delta\boldsymbol{x} \neq 0,\ \|G\Delta\boldsymbol{x}\|_1 > \langle \boldsymbol{w}, \Delta\boldsymbol{x}\rangle. \tag{11}$$

Let $H_w \subset \mathbb{R}^n$ be the affine subspace $\{\boldsymbol{x} \mid \langle \boldsymbol{w}, \boldsymbol{x}\rangle = 1\}$. The function $\|G \cdot\|_1$ defines a norm $\|\cdot\|_\diamond$ on $\mathbb{R}^n$. Geometrically, (11) is satisfied iff the unit ball $B_\diamond$ of $\|\cdot\|_\diamond$ is contained in the halfspace $H_w^- = \{\boldsymbol{x} \mid \langle \boldsymbol{w}, \boldsymbol{x}\rangle < 1\}$, as illustrated in Figure 3. This unit ball is a convex polytope, given by the inverse image (under the injective map $G$) of the intersection of $\mathcal{R}(G)$ and the unit $\ell^1$-ball $B_1$ in $\mathbb{R}^p$:

$$B_\diamond = G^{-1}[\, \mathcal{R}(G) \cap B_1(\mathbb{R}^p)\,]\,. \tag{12}$$

Now, $B_\diamond \subset H_w^-$ iff $[\mathcal{R}(G) \cap B_1(\mathbb{R}^p)] \subset G[\,H_w^-\,]$ iff $B_1(\mathbb{R}^p) \cap G[\,\mathrm{cl}\, H_w^+\,] = \emptyset$. These two closed convex sets are nonintersecting iff there is a hyperplane[8] $H_q = \{\boldsymbol{v} \in \mathbb{R}^p \mid \langle \boldsymbol{q}, \boldsymbol{v}\rangle = 1\} \subset \mathbb{R}^p$ separating them (see Figure 3 again). We lose no generality in assuming that $B_1 \subset H_q^-$, that $G[\mathrm{cl}\, H_w^+] \subset \mathrm{cl}\, H_q^+$, and that $H_q$ meets the relative boundary $\mathrm{rbd}\; G[\mathrm{cl}\, H_w^+] = G[H_w]$. The first condition occurs iff $\|\boldsymbol{q}\|_\infty < 1$, while the second occurs iff $G^*\boldsymbol{q} = \boldsymbol{w}$. ∎

The most natural candidate for a normal vector $\boldsymbol{q}$ is the minimum $\ell^2$-norm solution to this equation,

$$\boldsymbol{q}_0 = (G^\dagger)^* \boldsymbol{w} = G(G^*G)^{-1}\boldsymbol{w}. \tag{13}$$

[7]In the model outlined above, this occurs with probability one for $m$ sufficiently large.

[8]Notice $H_q$ cannot contain $0 \in \text{interior}(B_1)$, so the normalization $\langle \boldsymbol{q}, \boldsymbol{v}\rangle = 1$ is appropriate.

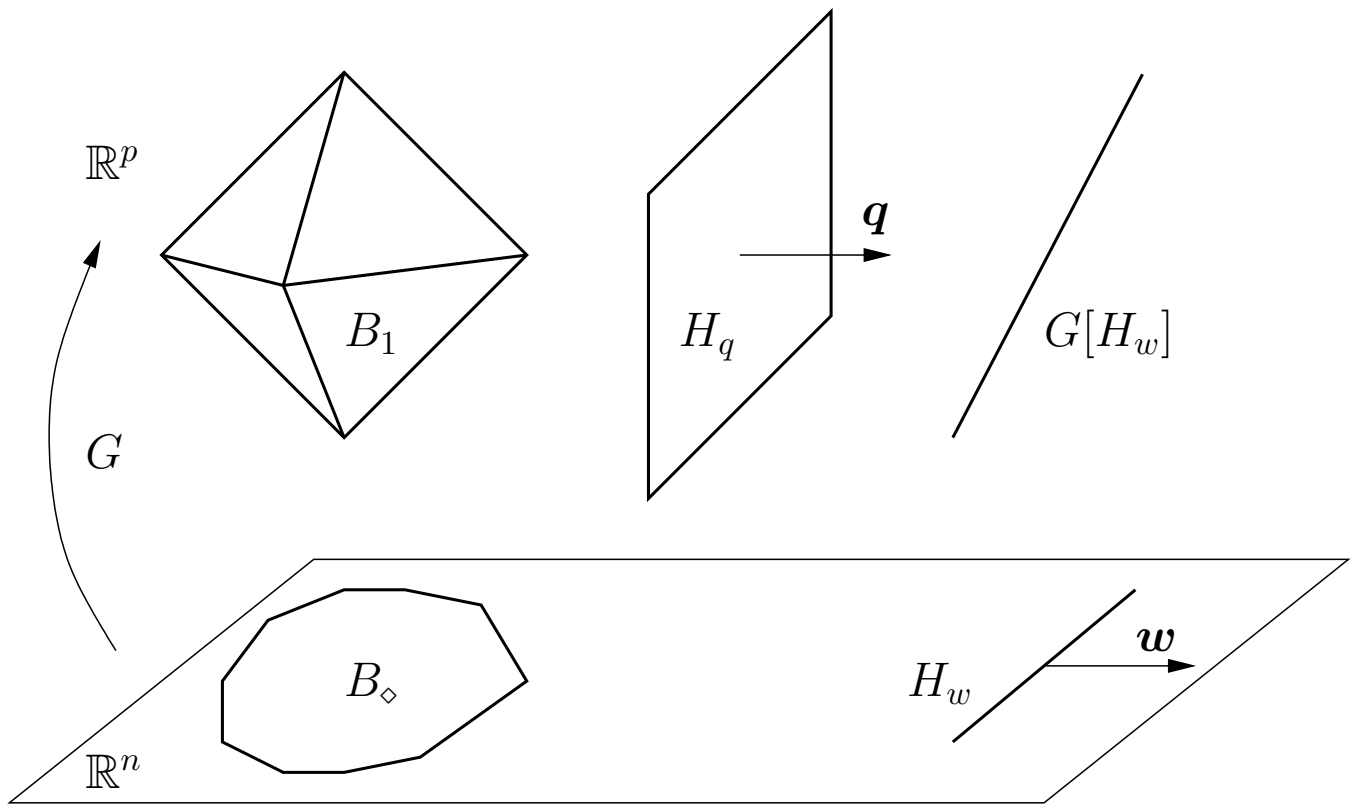

Fig. 3

**Geometry for the proof of Lemma 1.** THE UNIT BALL $B_\diamond$ CAN BE SEPARATED FROM $H_w$ IN $\mathbb{R}^n$ IF AND ONLY IF IN THE LIFTED SPACE $\mathbb{R}^p$, THE $\ell^1$-BALL $B_1$ CAN BE SEPARATED FROM THE IMAGE OF $H_w$ UNDER THE INJECTIVE MAP $G$. $H_q$ IS THE SEPARATING HYPERPLANE WITH A NORMAL VECTOR $\boldsymbol{q}$. SUCH AN $H_q$ MIGHT NOT BE UNIQUE IN $\mathbb{R}^p$, AND $\boldsymbol{q}_0$ WOULD BE THE NORMAL TO THE SPECIAL SEPARATING HYPERPLANE THAT CONTAINS $G(H_w)$.

When we use this particular normal $\boldsymbol{q}_0$, we are demanding that the *projection* of $B_1$ onto $\mathcal{R}(G)$ lie in $G[H_w^-]$. Since the projection contains the intersection, $B_1 \subset \{\langle \boldsymbol{q}_0, \cdot \rangle < 1\}$ is a sufficient, but not necessary condition. It is not surprising, then, that this condition often does not hold – empirically, $\|\boldsymbol{q}_0\|_\infty \geq 1$ with high probability. However, as we will see, the set of violations is almost always small, and we can apply a simple iterative scheme to improve $\boldsymbol{q}_0$ to a valid separator $\boldsymbol{q}$ with $\|\boldsymbol{q}\|_\infty < 1$.

## C. Iterative Construction of Separator

Our next lemma argues that if we are given an initial guess at a normal vector $\boldsymbol{q}_0 \in \mathbb{R}^p$ whose hyperplane $H_{q_0}$ separates $G[H_w]$ from *most* of the vertices of $B_1$, then we can refine $\boldsymbol{q}_0$ to a $\boldsymbol{q}_\infty$ that separates $G[H_w]$ and *all* of the vertices of $B_1$. In general, finding such a $\boldsymbol{q}_\infty$ requires solving a linear programming problem. We will analyze the feasibility of this linear program by considering an iteration similar to the alternating projection method for finding a pair of closest points between two convex sets. In this case, the two convex sets of interest are the hypercube of radius $1 - \varepsilon$ and the affine subspace $\boldsymbol{q}_0 + \mathcal{R}(G)^\perp$.

In the following lemma, $\boldsymbol{q}_0 \in \mathbb{R}^p$ is arbitrary (though $\boldsymbol{q}_0 = G^{\dagger *}\boldsymbol{w}$ is natural). We will construct a sequence of vectors $(\boldsymbol{q}_k)_{k=0}^\infty$. Fix a small constant $\varepsilon > 0$, and define the operator $\theta$ which takes the part of a vector that protrudes above $1-\varepsilon$:

$$[\theta \boldsymbol{x}](i) \doteq \begin{cases} 0, & \text{for } |\boldsymbol{x}(i)| \le 1-\varepsilon, \\ \operatorname{sgn}(\boldsymbol{x}(i))(|\boldsymbol{x}(i)|-1+\varepsilon), & \text{for } |\boldsymbol{x}(i)| > 1-\varepsilon. \end{cases} \tag{14}$$

We iteratively construct $\boldsymbol{q}_\infty$ by setting

$$\boldsymbol{q}_{i+1} \;=\; \boldsymbol{q}_i - \pi_{\mathcal{R}(G)^\perp}\theta \boldsymbol{q}_i \;=\; \boldsymbol{q}_i - \theta\boldsymbol{q}_i + \pi_{\mathcal{R}(G)}\theta\boldsymbol{q}_i. \tag{15}$$

Notice that by construction, $G^*\boldsymbol{q}_k = G^*\boldsymbol{q}_0 = \boldsymbol{w}$ for all $k$. So if $\theta\boldsymbol{q}_i \to 0$, then $\|\boldsymbol{q}_i\|_\infty < 1$ eventually, and $\boldsymbol{q}_\infty$ is a valid separator.

Before proving that this iteration produces a valid separator with high probability, we first demonstrate its behavior on a simulated example with $m = 3,000$, $\delta = .4$, $\nu = .1$, $\rho = .65$, and $k_1 = 10$. Figure 4 plots the sorted absolute values of entries of $\boldsymbol{q}_i$. Notice that the sorted coefficients clearly divide into two parts; these correspond to the upper[9] $(R_1)$ and lower $(R_2)$ indices. The initial separator $\boldsymbol{q}_0$ cleanly separates $G[H_w]$ from most of the vertices of $B_1$: only 39 entries protrude above $1-\varepsilon$. These entries are quickly iterated away: $\|\theta\boldsymbol{q}\|$ decreases geometrically until after 5 iterations a valid separator is obtained.

*Lemma 2:* Suppose $\exists\, c \in (0,1)$ such that

$$\xi \;\doteq\; \sup_{\|\boldsymbol{s}\|_0 \le cp,\ \boldsymbol{s}\neq\boldsymbol{0}} \frac{\|\pi_{\mathcal{R}(G)}\boldsymbol{s}\|_2}{\|\boldsymbol{s}\|_2} \;<\; 1, \tag{16}$$

and

$$\|\boldsymbol{q}_0\|_2 + \frac{1}{1-\xi}\|\theta\boldsymbol{q}_0\|_2 \le (1-\varepsilon)\sqrt{cp}, \tag{17}$$

where $G$ is the matrix defined in (8). Iteratively construct a sequence of vectors $\{\boldsymbol{q}_i\}$, with $\boldsymbol{q}_i = \boldsymbol{q}_{i-1} - \pi_{\mathcal{R}(G)^\perp}\theta\boldsymbol{q}_{i-1}$, where $\theta$ threshold-residual operator defined in (14). Then $\lim_{k\to\infty}\theta\boldsymbol{q}_k = 0$.

*Proof:* Let $T_k = \{\, i \mid |\boldsymbol{q}_k(i)| > 1-\varepsilon \,\} \subset [p]$, and consider the following three statements:

$$\|\boldsymbol{q}_k\|_2 \le \|\boldsymbol{q}_0\|_2 + \|\theta\boldsymbol{q}_0\|_2 \sum_{i=0}^{k} \xi^i, \qquad \|\theta\boldsymbol{q}_k\|_2 \le \|\theta\boldsymbol{q}_0\|_2\, \xi^k, \qquad \#T_k \le cp. \tag{18}$$

We will show by induction that these statements hold for all $k$, establishing the lemma. The first two statements of (18) hold trivially $k=0$. For $\#T_0$, notice that by (17),

$$\#T_0 \;\le\; \frac{\|\boldsymbol{q}_0\|_2^2}{(1-\varepsilon)^2} \;\le\; cp.$$

[9] Where necessary, we will use $R_1 = \{1,\dots,m-k_2\} \subset [p]$ to index the upper rows of $G$ (corresponding to $A$), and $R_2 = [p] \setminus R_1$ to index the lower rows.

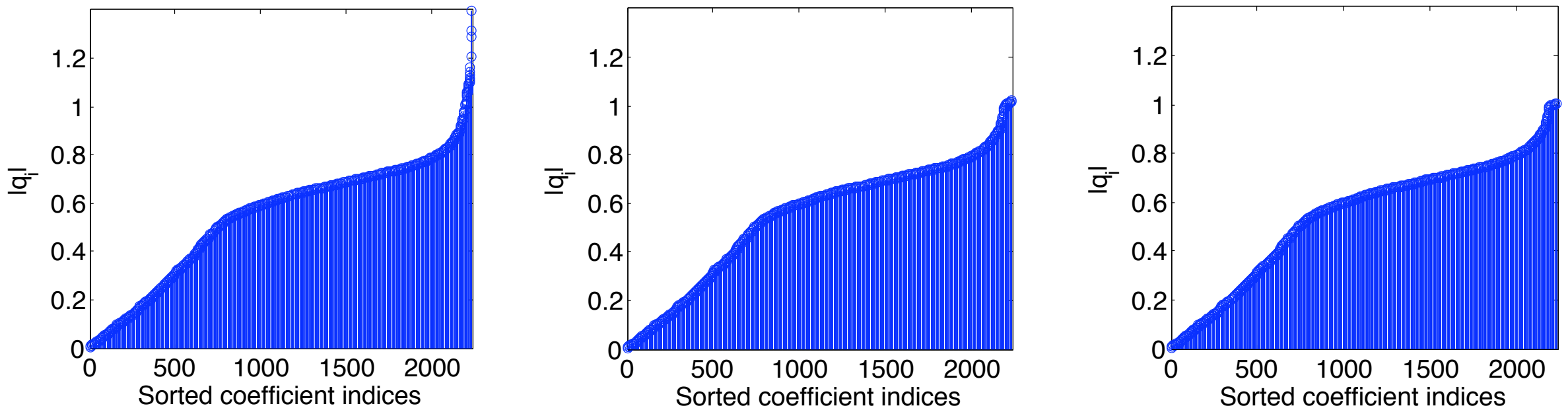

Iteration 0 $\#\{i \mid |\boldsymbol{q}_0(i)| > 1\} = 39,$ $\|\theta \boldsymbol{q}_0\|_2 = 1.176$

Iteration 1 $\#\{i \mid |\boldsymbol{q}_1(i)| > 1\} = 39$ $\|\theta \boldsymbol{q}_1\|_2 = 0.072$

Iteration 2 $\#\{i \mid |\boldsymbol{q}_2(i)| > 1\} = 38$ $\|\theta \boldsymbol{q}_2\|_2 = 0.006$

Fig. 4

**Iterative refinement producing a separating hyperplane.** HERE, $m = 3000$, $\delta = .4$, $\nu = .1$, $\rho = .65$, $k_1 = 10$. WE PLOT THE SORTED MAGNITUDES OF THE ENTRIES OF $\boldsymbol{q}_i$. AT LEFT, $\boldsymbol{q}_0$ SEPARATES $G(H_w)$ FROM MOST OF THE VERTICES OF $B_1$: ONLY 39 VIOLATIONS OCCUR. THE DISTINCT BIMODAL CHARACTERISTIC OF $\boldsymbol{q}_0$ IS DUE TO THE DIFFERENCES BETWEEN THE STATISTICS OF THE TOP ($R_1$) AND BOTTOM ($R_2$) INDICES. APPLYING THE ITERATION DECREASES $\|\theta \boldsymbol{q}_i\|$ GEOMETRICALLY; AFTER 5 ITERATIONS, A VALID SEPARATOR IS OBTAINED.

Now, suppose the three statements hold for $0, \dots, k$. Since $\theta \boldsymbol{q}_k$ has the same signs and smaller magnitude than $\boldsymbol{q}_k$, $\|\boldsymbol{q}_k - \theta \boldsymbol{q}_k\|_2 \le \|\boldsymbol{q}_k\|_2$; combining this with the inductive hypothesis we have

$$\begin{aligned} \|\boldsymbol{q}_{k+1}\|_2 &= \|\boldsymbol{q}_k - \theta \boldsymbol{q}_k + \pi_{\mathcal{R}(G)} \theta \boldsymbol{q}_k\| \le \|\boldsymbol{q}_k - \theta \boldsymbol{q}_k\| + \|\pi_{\mathcal{R}(G)} \theta \boldsymbol{q}_k\| \le \|\boldsymbol{q}_k\| + \xi^{k+1} \|\theta \boldsymbol{q}_0\| \\ &\le \|\boldsymbol{q}_0\|_2 + \|\theta \boldsymbol{q}_0\|_2 \sum_{i=0}^{k+1} \xi^i, \end{aligned}$$

Similarly, notice that since $\pi_{\mathcal{R}(G)} \theta \boldsymbol{q}_k$ dominates $\theta(\boldsymbol{q}_k - \theta \boldsymbol{q}_k + \pi_{\mathcal{R}(G)} \theta \boldsymbol{q}_k)$ elementwise,

$$\|\theta \boldsymbol{q}_{k+1}\| \;\le\; \|\pi_{\mathcal{R}(G)} \theta \boldsymbol{q}_k\| \;\le\; \xi \|\theta \boldsymbol{q}_k\| \;\le\; \xi^{k+1} \|\theta \boldsymbol{q}_0\|.$$

Finally, for the sparsity result $T_{k+1} \le cp$, note that

$$\|\boldsymbol{q}_{k+1}\|_2 \;\le\; \|\boldsymbol{q}_0\|_2 + \|\theta \boldsymbol{q}_0\| \sum_{i=0}^{k+1} \xi^i \;\le\; \|\boldsymbol{q}_0\|_2 + \frac{1}{1-\xi} \|\theta \boldsymbol{q}_0\|_2 \;\le\; (1-\varepsilon)\sqrt{cp},$$

and so $\theta \boldsymbol{q}_{k+1}$ must be $(cp)$-sparse. Since (18) holds for all $k$, $\|\theta \boldsymbol{q}_k\|_2 \to 0$. ■

### *D. Putting it All Together*

By Lemmas 1 and 2, if the two conditions (16) and (17) hold for a given sign and support triplet $(I, J, \boldsymbol{\sigma})$, then $(I, J, \boldsymbol{\sigma})$ is $\ell^1$-recoverable.[10] We will show that as $m \to \infty$, for any sequence of signal supports $I$, (16) and (17) hold with probability approaching one in the random matrix $A$ and error $(J, \boldsymbol{\sigma})$. The probability that either condition fails for a given $I$ will be small enough to allow a union bound over all $I$, establishing Theorem 1. We will assume we are in the large error regieme, with $\bar{\rho} \doteq 1 - \rho$ lower bounded as specified in the lemmas below. The conclusion still follows for smaller error fractions, since whenever $(I, J, \boldsymbol{\sigma})$ is $\ell^1$-recoverable, so is $(I, J', \boldsymbol{\sigma}_{J'})$ for any $J' \subset J$.

In this section, we lay out the main ideas for the rest of the proof, which consists of two parts, one for each of the conditions in Lemma 2. We establish that following two properties hold simultaneously with probability at least $1 - e^{-Cm^{1-\eta_0/2}(1+o(1))}$:

1) For a small enough constant $c$, the projection ratio $\xi$ for $cm$-sparse signals onto $\mathcal{R}(G)$ is bounded below 1 by a polynomial function in $\nu$. More precisely, $\xi < 1 - C\nu^8$ for some constant $C > 0$. As a result, the coefficient $\frac{1}{1-\xi}$ in the second condition (17) is bounded by $C^{-1}\nu^{-8}$.
2) As $m$ goes to infinity, the $\ell^2$-norm of the initial separating normal vector $\|\boldsymbol{q}_0\|_2$ is bounded above by $\nu O(m^{1/2})$, and $\|\theta \boldsymbol{q}_0\|_2$ is bounded above by $e^{-\alpha/\nu^2} O(m^{1/2})$ for some constant $\alpha$.

Putting these results together, the initial separating normal vector $\boldsymbol{q}_0$ satisfies:

$$\|\boldsymbol{q}_0\|_2 + \frac{1}{1-\xi}\|\theta \boldsymbol{q}_0\|_2 \;\leq\; \nu O(m^{1/2}) + C^{-1}\nu^{-8} e^{-\alpha/\nu^2} O(m^{1/2}). \tag{19}$$

If the deviation $\nu$ of the bouquet is small enough, the second condition (17) of Lemma 2 will be satisfied, since the right hand side, $(1-\varepsilon)\sqrt{cp} = \Omega(m^{1/2})$ is independent of $\nu$. Hence, by Lemma 2, the initial normal $\boldsymbol{q}_0$ will converge to a valid normal vector that separates the $\ell^1$-ball $B_1$ from the subspace $G[\,H_w\,]$, establishing $\ell^1$-recoverability at $(I, J, \boldsymbol{\sigma})$. Comparing the failure probability for the two conditions to the number of subsets $I \subset [n]$ of size $C_0 m^{1-\eta_0}$ then completes the proof of Theorem 1. These arguments are laid out more precisely and quantitatively in Section C of the appendix.

Whereas Lemmas 1 and 2 have simple geometric and algebraic proofs, the above results require more detailed analysis of large Gaussian matrices. We outline the main ideas of their proof in this section, leaving many of the technical details to the appendix. The derivation is based on recent (and now widely-used) results on concentration of Lipschitz functions [3], which state that if $\boldsymbol{x}$ is a $d$-dimensional

[10] Notice that conditions (16) and (17) depend on $(I, J, \boldsymbol{\sigma})$, through the construction of the matrix $G$.

iid $\mathcal{N}(0,1)$ vector and $f : \mathbb{R}^d \to \mathbb{R}$ is 1-Lipschitz, then

$$P\left[\,|f(\boldsymbol{x}) - \mathbb{E}f(\boldsymbol{x})| \geq t\,\right] \;\leq\; 2\exp\left(-\frac{2t^2}{\pi^2}\right). \tag{20}$$

Two cases are of particular interest here. First, the norm concentrates as (see, e.g., [27]):

$$P\left[\,\|\boldsymbol{x}\|_2 \;\geq\; \beta\sqrt{d}\right] \;\leq\; \exp\left(-\frac{2(\beta-1)^2}{\pi^2}\,d\right). \tag{21}$$

Second, as has been widely exploited in the compressed sensing literature (e.g., [14], [16]), the singular values of rectangular Gaussian matrices with aspect ratio $\alpha$ concentrate about the values $1 \pm \sqrt{\alpha}$ predicted by the Marchenko-Pasteur law:

*Fact 1 (Concentration of singular values [3]):* Let $A \in \mathbb{R}^{m\times n}$, $(m > n)$ be a random matrix with entries iid $\mathcal{N}(0, \frac{1}{m})$. Then for any $t > 0$,

$$P\left[\sigma_{max}(A) > 1 + \sqrt{n/m} + o(1) + t\right] \quad\leq\quad e^{-mt^2/2}, \tag{22}$$

$$P\left[\sigma_{min}(A) < 1 - \sqrt{n/m} + o(1) - t\right] \quad\leq\quad e^{-mt^2/2}. \tag{23}$$

We will also return to (20) in the proof of Lemma 8 of the appendix.

*1) Projection of Sparse Vectors:* In this subsection, we upper bound the norm of the projection of any sparse vector onto $\mathcal{R}(G)$. Since the lower $(R_2)$ coordinates of

$$G \doteq \begin{bmatrix} A_1 & A_2 \\ 0 & \mathtt{I} \end{bmatrix} = \begin{bmatrix} Z_1 + \boldsymbol{\mu}_{J^c}\mathbf{1}_{k_1}^* & Z_2 + \boldsymbol{\mu}_{J^c}\mathbf{1}_{\delta m-k_1}^* \\ 0 & \mathtt{I} \end{bmatrix}$$

contain an identity matrix, when the variance $\nu^2/m$ of the perturbations $Z_1, Z_2$ is small, we expect that sparse vectors with support on $R_2$ will be very close to $\mathcal{R}(G)$. The following lemma verifies that this is the case, but argues that distance to $\mathcal{R}(G)$ is at least $\Omega(\nu^8)$. The technical conditions appear complicated, but simply assert that the fraction of nonzeros $c$ is sufficiently small.

*Lemma 3 (Projection of Sparse Vectors):* Suppose that $\bar\rho < \delta$ and $\nu < \min\left(\frac{1}{9}, (512/\delta)^{1/4}\right)$,

$$c \;<\; \min\left\{\frac{\bar\rho}{1024}, \frac{\bar\rho}{64(1+2C_\mu\,\bar\rho^{-1/2})^2}\right\}, \quad \bar\rho H(c/\bar\rho) + \delta H(c/\delta) < \frac{\bar\rho}{128\pi^2}, \tag{24}$$

where $H(\cdot)$ is the base-$e$ binary entropy function. Then the projection of a sparse vector $\boldsymbol{s} \in \mathbb{R}^p$ with $\|\boldsymbol{s}\|_0 \leq cm$ onto the range of $G$ is bounded as

$$\sup_{\|\boldsymbol{s}\|_0 \leq cm,\; \boldsymbol{s}\neq \mathbf{0}} \frac{\|\pi_{\mathcal{R}(G)}\boldsymbol{s}\|_2}{\|\boldsymbol{s}\|_2} \;<\; 1 - \nu^8\left(\frac{\sqrt{\bar\rho}\,(\sqrt{\delta}-\sqrt{\bar\rho})}{32 + 128\,\nu^2\,(\sqrt{\delta}+\sqrt{\bar\rho})^2}\right)^4 \tag{25}$$

on the complement of a bad event with probability $e^{-Cm\,(1+o(1))}$.

*Proof:* The projection of $\boldsymbol{s} = \left[\begin{smallmatrix}\boldsymbol{s}_1\\ \boldsymbol{s}_2\end{smallmatrix}\right]$ onto $\mathcal{R}(G)$ solves

$$\min_{\boldsymbol{r}\in\mathbb{R}^n} \left\|\left[\begin{smallmatrix}\boldsymbol{s}_1\\ \boldsymbol{s}_2\end{smallmatrix}\right] - G\boldsymbol{r}\right\|_2^2 \;=\; \min_{\boldsymbol{u}_1,\boldsymbol{u}_2} \left\|\left[\begin{smallmatrix}\boldsymbol{s}_1\\ \boldsymbol{s}_2\end{smallmatrix}\right] - G\left[\begin{smallmatrix}\boldsymbol{u}_1\\ \boldsymbol{s}_2+\boldsymbol{u}_2\end{smallmatrix}\right]\right\|_2^2 \;=\; \min_{\boldsymbol{u}_1,\boldsymbol{u}_2} \|\boldsymbol{s}_1 - A_1\boldsymbol{u}_1 - A_2(\boldsymbol{s}_2+\boldsymbol{u}_2)\|_2^2 + \boldsymbol{u}_2^*\boldsymbol{u}_2.$$

By minimizing the first term, we can write the unique optimal $\boldsymbol{u}_1$ in terms of the remaining variables:

$$\boldsymbol{u}_1 \;=\; (A_1^* A_1)^{-1} A_1^* \boldsymbol{s}_1 - (A_1^* A_1)^{-1} A_1^* A_2 (\boldsymbol{s}_2 + \boldsymbol{u}_2)$$

and subsequently, the optimal $\boldsymbol{u}_2$ satisfies:

$$-A_2^* \boldsymbol{s}_1 + A_2^* A_1 \boldsymbol{u}_1 + A_2^* A_2 (\boldsymbol{s}_2 + \boldsymbol{u}_2) + \boldsymbol{u}_2 = \mathbf{0} \;\Rightarrow\; \left(\mathtt{I} + A_2^* \pi_{A_1^\perp} A_2\right) \boldsymbol{u}_2 = A_2^* \pi_{A_1^\perp} \boldsymbol{s}_1 - A_2^* \pi_{A_1^\perp} A_2 \boldsymbol{s}_2,$$

where $\pi_{A_1^\perp}$ denotes the projection matrix onto the orthogonal complement of $\mathcal{R}(A_1)$.

Write $A_2^* \pi_{A_1^\perp} = U S V^*$ with $U \in \mathbb{R}^{(\delta m - k_1) \times (\bar\rho m - k_1)}$ and $V \in \mathbb{R}^{\bar\rho m \times (\bar\rho m - k_1)}$ orthogonal matrices, and the diagonal of $S \in \mathbb{R}^{(\bar\rho m - k_1) \times (\bar\rho m - k_1)}$ containing the nonzero singular values of $A_2^* \pi_{A_1^\perp}$. Then if $\boldsymbol{u}_2$ is the solution to the above equation

$$\begin{aligned} \|\, \boldsymbol{s} - \pi_{\mathcal{R}(G)} \boldsymbol{s} \,\|_2 \;\;\geq\;\; \|\boldsymbol{u}_2\|_2 \;&=\; \left\| (S^2 + \mathtt{I})^{-1} S V^* [\, \mathtt{I} \;\; - A_2 \,] \left[\begin{smallmatrix} \boldsymbol{s}_1 \\ \boldsymbol{s}_2 \end{smallmatrix}\right] \right\|_2 \\ &=\; \left\| (S^2 + \mathtt{I})^{-1} S \,[\, V^* \;\; - S U^* \,] \left[\begin{smallmatrix} \boldsymbol{s}_1 \\ \boldsymbol{s}_2 \end{smallmatrix}\right] \right\|_2 . \end{aligned} \tag{26}$$

Above is the norm of the product of a diagonal matrix $(S^2 + \mathtt{I})^{-1} S$, a wide matrix $[\, V^* \;\; - S U^* \,]$, and a sparse vector $\boldsymbol{s}$. We will bound it by lower bounding the elements of the diagonal matrix, and then lower bounding the "restricted minimum singular value"

$$\gamma_{cm}(\, [\, V^* \;\; - S U^* \,] \,) \;\;\doteq\;\; \inf_{\|\boldsymbol{s}\|_0 \,\leq\, cp,\; \boldsymbol{s} \neq \mathbf{0}} \frac{\| [\, V^* \;\; - S U^* \,] \, \boldsymbol{s} \,\|_2}{\|\boldsymbol{s}\|_2}.$$

We first drop the top row of $(S^2 + \mathtt{I})^{-1} S \,[\, V^* \;\; - S U^* \,]$. This allows us to uniformly lower bound the diagonal of $(S^2 + \mathtt{I})^{-1} S$. While $\sigma_1$ can be quite large due to the inhomogeneous term $(\boldsymbol{\mu}_{J^c} \mathbf{1}^*)$, and hence $\frac{\sigma_1}{\sigma_1^2 + 1}$ can be quite small, for the remaining singular values $\frac{\sigma_i}{\sigma_i^2 + 1}$ is at least on the order of $\nu$. Let $\tilde S \in \mathbb{R}^{(\bar\rho m - k_1 - 1) \times (\bar\rho m - k_1 - 1)}$ be the diagonal matrix obtained by dropping the row and column of $S$ corresponding to the largest singular value; $\tilde V$ and $\tilde U$ are obtained by dropping the corresponding columns. From (26),

$$\|\boldsymbol{u}_2\|_2 \;\;\geq\;\; \left\| (\tilde S^2 + \mathtt{I})^{-1} \tilde S \,[\, \tilde V^* \;\; - \tilde S \tilde U^* \,] \left[\begin{smallmatrix} \boldsymbol{s}_1 \\ \boldsymbol{s}_2 \end{smallmatrix}\right] \right\|_2 \;\geq\; \frac{\sigma_{min}(A_2^* \pi_{A_1^\perp})}{1 + \sigma_2^2(A_2^* \pi_{A_1^\perp})} \gamma_{cm}([\, \tilde V^* \;\; - \tilde S \tilde U^* \,]) \, \|\boldsymbol{s}\|_2, \tag{27}$$

where $\sigma_{min}(A_2^* \pi_{A_1^\perp})$ is the smallest nonzero singular value and $\sigma_2(A_2^* \pi_{A_1^\perp})$ is the second largest singular value.

*a) Bounding the second largest singular value $\sigma_2(A_2^* \pi_{A_1^\perp})$:* Write $\hat{\boldsymbol{\mu}} \doteq \pi_{A_1^\perp} \boldsymbol{\mu}_{J^c}$, and notice that

$$\begin{aligned} \sigma_2(A_2^* \pi_{A_1^\perp}) \;&=\; \inf_{\boldsymbol{u} \neq \mathbf{0}} \sup_{\boldsymbol{v} \neq \mathbf{0}} \frac{\|A_2^* \pi_{A_1^\perp} \pi_{\boldsymbol{u}^\perp} \boldsymbol{v}\|_2}{\|\boldsymbol{v}\|_2} \;=\; \inf_{\boldsymbol{u} \neq \mathbf{0}} \sigma_1(A_2^* \pi_{A_1^\perp} \pi_{\boldsymbol{u}^\perp}) \\ &\leq\; \sigma_1(A_2^* \pi_{A_1^\perp} \pi_{\hat{\boldsymbol{\mu}}^\perp}) \;=\; \sigma_1(Z_2^* \pi_{(\boldsymbol{\mu}_{J^c}, Z_1)^\perp}). \end{aligned}$$

Choose any orthonormal basis for the subspace $\Sigma = (\mathcal{R}(Z_1) + \mathcal{R}(\boldsymbol{\mu}_{J^c}))^{\perp}$. Since $\Sigma$ is probabilistically independent of $Z_2$, the representation of the projection $Z_2^* \pi_\Sigma$ with respect to the chosen basis is simply distributed as a $(\delta m - k_1) \times (\bar{\rho} m - k_1 - 1)$ random matrix $\hat{Z}_2$ with entries $\mathcal{N}(0, \nu^2/m)$. Since $\frac{\sqrt{m}}{\nu\sqrt{\delta m - k_1}} \hat{Z}_2$ is $\mathcal{N}(0, \frac{1}{\delta m - k_1})$, by Fact 1,

$$P\left[\sigma_1\left(\frac{\sqrt{m}}{\nu\sqrt{\delta m - k_1}} \hat{Z}_2\right) \geq 1 + \sqrt{\frac{\bar{\rho} m - k_1 - 1}{\delta m - k_1}} + t\right] \;\leq\; \exp\left(-(t - o(1))^2 (\delta m - k_1)/2\right), \tag{28}$$

and so $P\left[\sigma_1(\hat{Z}_2) \geq 2\nu(\sqrt{\delta} + \sqrt{\bar{\rho}})\right] \leq e^{-Cm\,(1+o(1))}$. On the complement of this bad event, $\sigma_2^2(A_2^* \pi_{A_1^\perp}) \leq 4\nu^2(\sqrt{\delta} + \sqrt{\bar{\rho}})^2$.

*b) Bounding the smallest nonzero singular value* $\sigma_{min}(A_2^* \pi_{A_1^\perp}) = \inf_{\boldsymbol{x} \in A_1^\perp} \frac{\|A_2^* \boldsymbol{x}\|_2}{\|\boldsymbol{x}\|_2}$*:* Let $W \in \mathbb{R}^{\bar{\rho} m \times (\bar{\rho} m - k_1)}$ be a matrix whose columns form an orthonormal basis for $A_1^\perp$, and let $Q \in \mathbb{R}^{(\delta m - k_1) \times (\delta m - k_1 - 1)}$ be an orthonormal basis for $\mathbf{1}_{\delta m - k_1}^\perp$. Then $\sigma_{min}(A_2^* \pi_{A_1^\perp}) \geq \sigma_{min}(Q^* A_2 W)$. Conditioned on $A_1$, $Z_2' \doteq Q^* A_2 W \in \mathbb{R}^{(\delta m - k_1 - 1) \times (\bar{\rho} m - k_1)}$ is iid $\mathcal{N}(0, \nu^2/m)$. Applying Fact 1 (with a similar rescaling argument to the one used for $\sigma_{max}(\hat{Z}_2)$ above) gives that

$$P\left[\sigma_{min}(Z_2') < \frac{\nu}{2}\left(\sqrt{\delta} - \sqrt{\bar{\rho}}\right)\right] \leq e^{-Cm(1+o(1))}. \tag{29}$$

On the complement of this bad event, $\sigma_{min}(A_2^* \pi_{A_1^\perp}) \geq \frac{\nu}{2}(\sqrt{\delta} - \sqrt{\bar{\rho}})$.

Finally, in Lemma 5 of Appendix A, we show that under the stated conditions, the restricted singular value $\gamma_{cm}$ in (27) satisfies $\gamma_{cm}(\,[\tilde{V}^* \quad -\tilde{S}\tilde{U}^*]\,) \geq \frac{\nu\sqrt{\bar{\rho}}}{16}$ with probability at least $1 - e^{-Cm(1+o(1))}$. Notice that this bound agrees with (and in fact is looser than) the Marchenko-Pasteur law for a $\bar{\rho} m \times cm$ Gaussian $\mathcal{N}(0, \nu^2/m)$ matrix (i.e., the concentration result of Fact 1). In fact, the proof argues that the two blocks of this matrix are probabilistically independent, and then applies Fact 1 to an equivalent pair of Gaussian matrices. The somewhat technical conditions (24) introduced here are necessary to ensure that a union bound over all subsets of $cm$ columns remains small.

Combining the three results, we have that for all $\boldsymbol{s} \in \mathbb{R}^p$ with $\|\boldsymbol{s}\|_0 \leq cm$,

$$\frac{\|\boldsymbol{s} - \pi_G \boldsymbol{s}\|_2}{\|\boldsymbol{s}\|_2} \;\geq\; \frac{\nu^2\sqrt{\bar{\rho}}\,(\sqrt{\delta} - \sqrt{\bar{\rho}}\,)}{32 + 128\,\nu^2\,(\sqrt{\bar{\rho}} + \sqrt{\delta})^2} \;\doteq\; \beta \tag{30}$$

Notice that $\frac{\|\pi_G \boldsymbol{s}\|}{\|\boldsymbol{s}\|} = \sqrt{1 - \left(\frac{\|\boldsymbol{s} - \pi_G \boldsymbol{s}\|}{\|\boldsymbol{s}\|}\right)^2} \leq \sqrt{1 - \beta^2} \leq 1 - \beta^4$, where we have used that $1 - \beta^4 > \sqrt{1 - \beta^2}$ for $\beta < 1/\sqrt{2}$; this is guaranteed for $\nu < (512/\delta)^{1/4}$. Combined with (30), this implies (25). ■

*2) Initial Separating Hyperplane:* In this section, we analyze the initial separator $\boldsymbol{q}_0$, obtained as the minimum 2-norm solution to the equation $G^* \boldsymbol{q} = \boldsymbol{w}$. We upper bound both $\|\boldsymbol{q}_0\|_2$ and $\|\theta \boldsymbol{q}_0\|_2$, where the operator $\theta$ defined in (14) retains the portion of a vector that protrudes above $1 - \varepsilon$ in absolute value.

These bounds provide the second half of the conditions needed in Lemma 2 to show that $\boldsymbol{q}_0$ can be refined by alternating projections to give a true separator.

*Lemma 4:* Suppose $\bar{\rho} < \delta$ and $\nu < \frac{1}{8\,(\sqrt{\delta}+1)}$. Then for $G$ defined in (8) and $\boldsymbol{w} = A_{J,\bullet}^{*}\boldsymbol{\sigma} - \mathbf{1}_I$, $\exists$ constants $\alpha_1$, $\alpha_2$ such that $\boldsymbol{q}_0 = G^{\dagger *}\boldsymbol{w}$ satisfies

$$\|\boldsymbol{q}_0\|_2 \;\leq\; \alpha_1\,\nu\, m^{1/2} \;+\; o(m^{1/2}), \tag{31}$$

$$\|\theta\boldsymbol{q}_0\|_2 \;\leq\; \alpha_2 \exp\left(-\frac{1}{64\nu^2}\right) m^{1/2} \;+\; o(m^{1/2}). \tag{32}$$

on the complement of a bad event of probability $\leq e^{-Cm^{1-\eta_0/2}(1+o(1))}$.

*Proof:* Notice that $G^{\dagger *} = G(G^*G)^{-1} = \left[\begin{smallmatrix} Z_1 & Z_2 \\ 0 & \mathrm{I} \end{smallmatrix}\right](G^*G)^{-1} + \left[\begin{smallmatrix} \boldsymbol{\mu}_{J^c}\mathbf{1}^* \\ 0 \end{smallmatrix}\right](G^*G)^{-1}$, where $Z_1 \doteq Z_{J^c,I}$ and $Z_2 \doteq Z_{J^c,I^c}$. Expanding $\boldsymbol{q}_0 = G^{\dagger *}\boldsymbol{w}$ gives

$$\begin{aligned} \boldsymbol{q}_0 \;=\;& \left[\begin{smallmatrix} Z_1 & Z_2 \\ 0 & \mathrm{I} \end{smallmatrix}\right](G^*G)^{-1}Z_{J,\bullet}^*\boldsymbol{\sigma} \;+\; \left[\begin{smallmatrix} Z_1 & Z_2 \\ 0 & \mathrm{I} \end{smallmatrix}\right]\Big(-(G^*G)^{-1}\mathbf{1}_I + \langle \boldsymbol{\mu}_J, \boldsymbol{\sigma}\rangle\,(G^*G)^{-1}\mathbf{1}\Big) \\ &+\; \left[\begin{smallmatrix} \boldsymbol{\mu}_{J^c} \\ \mathbf{0} \end{smallmatrix}\right]\Big(\mathbf{1}^*(G^*G)^{-1}Z_{J,\bullet}^*\boldsymbol{\sigma} \;-\; \mathbf{1}^*(G^*G)^{-1}\mathbf{1}_I \;+\; \langle \boldsymbol{\mu}_J, \boldsymbol{\sigma}\rangle\,\mathbf{1}^*(G^*G)^{-1}\mathbf{1}\Big). \end{aligned} \tag{33}$$

In this section, we concentrate our efforts on the first term above. In Lemma 7 of Appendix B, we give a more detailed analysis of $(G^*G)^{-1}$, which shows that the remaining terms are all negligible, contributing $o(m^{1/2})$ to $\|\boldsymbol{q}_0\|$. This is essentially due to the presence of a large common term $\boldsymbol{\mu}_{J^c}$ in the columns of $G$: the most significant term in $G^*G$ is $\boldsymbol{\mu}_{J^c}^*\boldsymbol{\mu}_{J^c}\mathbf{1}\mathbf{1}^*$, and $(G^*G)^{-1}$ shrinks $\mathbf{1}$. More precisely, Lemma 7 of Appendix B shows that with probability at least $1 - e^{-Cm^{1-\eta_0/2}(1+o(1))}$,

$$\left\| \boldsymbol{q}_0 - \left[\begin{smallmatrix} Z_1 & Z_2 \\ 0 & \mathrm{I} \end{smallmatrix}\right](G^*G)^{-1}Z_{J,\bullet}^*\boldsymbol{\sigma} \right\| \;\leq\; Cm^{1/2-\eta_0/4}.$$

This remaining term can be further simplified by splitting out several of the inhomogeneous parts of $(G^*G)^{-1}$. Define $Q \doteq Z_{J^c,\bullet}^*Z_{J^c,\bullet} + \left[\begin{smallmatrix} 0 & 0 \\ 0 & \mathrm{I} \end{smallmatrix}\right] = \left[\begin{smallmatrix} Z_1^*Z_1 & Z_1^*Z_2 \\ Z_2^*Z_1 & Z_2^*Z_2+\mathrm{I} \end{smallmatrix}\right] \in \mathbb{R}^{n\times n}$ and $\boldsymbol{\zeta} \doteq Z_{J^c,\bullet}^*\boldsymbol{\mu}_{J^c} \in \mathbb{R}^n$. In terms of these variables, $G^*G = Q + \boldsymbol{\zeta}\mathbf{1}^* + \mathbf{1}\boldsymbol{\zeta}^* + \alpha\mathbf{1}\mathbf{1}^*$. Applying the matrix inversion lemma,

$$(G^*G)^{-1} = Q^{-1} - Q^{-1/2}M\Xi M^*Q^{-1/2}, \tag{34}$$

where $M = \left[\begin{array}{cc} \frac{Q^{-1/2}\mathbf{1}}{\|Q^{-1/2}\mathbf{1}\|_2} & \frac{Q^{-1/2}\boldsymbol{\zeta}}{\|Q^{-1/2}\boldsymbol{\zeta}\|_2} \end{array}\right] \in \mathbb{R}^{n\times 2}$, and $\Xi$ is an appropriate $2\times 2$ matrix. Since $\boldsymbol{\vartheta} \doteq Z_{J,\bullet}^*\boldsymbol{\sigma} \in \mathbb{R}^n$ is iid $\mathcal{N}(0,\nu^2\rho)$ independent of $G$, with high probability it is almost orthogonal to the rank-2 perturbation $\Gamma \doteq Q^{-1/2}M\Xi M^*Q^{-1/2}$: $P\left[\|\pi_\Gamma\boldsymbol{\vartheta}\| \geq m^{1/2-\eta_0/4}\right] \asymp e^{-Cm^{1-\eta_0/2}}$.[11] Using Fact 1 and block singular value identities, it is not difficult to show[12] that $\|Q^{-1}\| \leq \frac{4}{\nu^2\bar{\rho}}$ with probability at

[11] $\|\pi_\Gamma\boldsymbol{\vartheta}\|$ is distributed as the norm of a 2-dimensional $\mathcal{N}(0,\nu^2\rho)$ vector. The bound follows from the $\chi$ tail bound (21).

[12] Use that $\sigma_{min}^2\left(\left[\begin{smallmatrix} Z_1 & Z_2 \\ 0 & \mathrm{I} \end{smallmatrix}\right]\right) \geq \sigma_{min}^2(Z_1) - \frac{\|Z_1\|^2\|Z_2\|^2}{1-\sigma_{min}^2(Z_1)}$ and apply Fact 1 to bound each term.

least $1 - e^{-Cm(1+o(1))}$. Combined with the bound $\|(G^*G)^{-1}\| \le C_G$ from Lemma 7, we have that $\|\Gamma\| \le \|(G^*G)^{-1}\| + \|Q^{-1}\| \le C_G + \frac{4}{\nu^2\bar\rho}$ is bounded by a constant, and

$$\left\|\left[\begin{smallmatrix} Z_1 & Z_2 \\ 0 & \mathtt{I}\end{smallmatrix}\right]\Gamma\boldsymbol{\vartheta}\right\| \;\le\; \left\|\left[\begin{smallmatrix} Z_1 & Z_2 \\ 0 & \mathtt{I}\end{smallmatrix}\right]\right\| \, \|\Gamma\| \, \|\pi_\Gamma\boldsymbol{\vartheta}\| \;\le\; \left(1 + 2\nu^2(\sqrt{\rho}+\sqrt{\delta})^2\right)^{1/2}\left(C_G + \frac{4}{\nu^2\bar\rho}\right) m^{1/2-\eta_0/4}$$

and the remaining part of $\boldsymbol{q}_0$ is

$$\left[\begin{smallmatrix} Z_1 & Z_2 \\ 0 & \mathtt{I}\end{smallmatrix}\right] Q^{-1}\boldsymbol{\vartheta} \;=\; \left[\begin{smallmatrix} Z_1 \\ 0 \end{smallmatrix}\right]\left[Q^{-1}\right]_{I,\bullet}\boldsymbol{\vartheta} \;+\; \left[\begin{smallmatrix} Z_2 \\ \mathtt{I} \end{smallmatrix}\right]\left[Q^{-1}\right]_{I^c,I}\boldsymbol{\vartheta}_I \;+\; \left[\begin{smallmatrix} Z_2 \\ \mathtt{I} \end{smallmatrix}\right]\left[Q^{-1}\right]_{I^c,I^c}\boldsymbol{\vartheta}_{I^c}.$$

The first two terms involve projections of $\boldsymbol{\vartheta}$ onto $k_1$-dimensional subspaces, and hence are of lower order. That is, for $\Sigma \doteq \mathrm{null}([Q^{-1}]_{I,\bullet})^\perp$, we have $P\left[\|\pi_\Sigma\boldsymbol{\vartheta}\|_2 \ge m^{1/2-\eta_0/4}\right] \asymp e^{-Cm^{1-\eta_0/2}}$. Since $\|Z_1\|$ and $\|Q^{-1}\|$ are bounded by constants with overwhelming probability, with probability at least $1 - e^{-Cm^{1-\eta_0/2}(1+o(1))}$, $\left\|\left[\begin{smallmatrix} Z_1 \\ 0 \end{smallmatrix}\right]\left[Q^{-1}\right]_{I,\bullet}\boldsymbol{\vartheta}\right\| \le C' m^{1/2-\eta_0/4}$. Identical reasoning shows that on the complement of a bad event of probability $\asymp e^{-Cm^{1-\eta_0/2}}$, $\left\|\left[\begin{smallmatrix} Z_2 \\ \mathtt{I} \end{smallmatrix}\right]\left[Q^{-1}\right]_{I^c,I}\boldsymbol{\vartheta}_I\right\| \le C'' m^{1/2-\eta_0/4}$.

This leaves $\left[\begin{smallmatrix} Z_2 \\ \mathtt{I} \end{smallmatrix}\right]\left[Q^{-1}\right]_{I^c,I^c}\boldsymbol{\vartheta}_{I^c}$. Expressing $Q$ as $\left[\begin{smallmatrix} U & V^* \\ V & W \end{smallmatrix}\right]$ and applying the Schur complement formula gives $[Q^{-1}]_{I^c,I^c} = W^{-1} + W^{-1}V(U^{-1} - V^*W^{-1}V)^{-1}V^*W^{-1}$, where $W = Z_2^*Z_2 + \mathtt{I}$, $V = Z_2^*Z_1$, and $U = Z_1^*Z_1$. Because $W \succeq \mathtt{I}$, $\|W^{-1}\| \le 1$. With probability at least $1 - e^{-Cm(1+o(1))}$, $\|U\| = \|Z_1\|^2 \le 2\nu^2\bar\rho$, $\sigma_{min}(U) \ge \frac{\nu^2\bar\rho}{2}$, and $\|V\| \le \|Z_1\|\|Z_2\| \le 2\,\nu^2\,(\sqrt{\bar\rho\delta} + \bar\rho)$ and so

$$\left\|W^{-1}V(U^{-1} - V^*W^{-1}V)^{-1}V^*W^{-1}\right\| \;\le\; \frac{\|W^{-1}\|^2\|V\|^2}{\sigma_{min}(U^{-1}) - \|V\|^2\|W^{-1}\|} \;\le\; \frac{8\nu^6(1+\sqrt{\delta})^2}{1 - 8\nu^6(1+\sqrt{\delta})^2}$$

is bounded by a constant. Let $\Sigma'$ denote the $k_1$-dimensional range of this matrix. With probability $\ge 1 - e^{-Cm^{1-\eta_0/2}(1+o(1))}$, $\|\pi_{\Sigma'}\boldsymbol{\vartheta}\| \le m^{1/2-\eta_0/4}$, and so

$$\left\|\left[\begin{smallmatrix} Z_2 \\ \mathtt{I} \end{smallmatrix}\right] W^{-1}V(U^{-1} - V^*W^{-1}V)^{-1}V^*W^{-1}\boldsymbol{\vartheta}\right\| \;\le\; C'''\, m^{1/2-\eta_0/4},$$

leaving only $\hat{\boldsymbol{q}}_0 \doteq \left[\begin{smallmatrix} Z_2 \\ \mathtt{I} \end{smallmatrix}\right](Z_2^*Z_2 + \mathtt{I})^{-1}\boldsymbol{\vartheta}_{I^c}$. With probability at least $1 - e^{-Cm(1+o(1))}$, $\|\boldsymbol{\vartheta}_{I^c}\| \le \sqrt{2}\,\nu\sqrt{\rho\delta}\, m^{1/2}$, and so

$$\|\hat{\boldsymbol{q}}_0\|_2 \;\le\; \left\|\left[\begin{smallmatrix} Z_2 \\ \mathtt{I} \end{smallmatrix}\right]\right\| \, \|\boldsymbol{\vartheta}_{I^c}\| \;\le\; \sqrt{1 + \|Z_2\|_2^2}\, \|\boldsymbol{\vartheta}_{I^c}\| \;\le\; \nu\sqrt{2\,\delta\,\rho\left(1 + 2\nu^2\left(\sqrt{\delta} + \sqrt{\bar\rho}\right)^2\right)}\, m^{1/2} \tag{35}$$

establishing the first part of the lemma.

For the second part, we will show that the the upper ($R_1$) and lower ($R_2$) parts of $\hat{\boldsymbol{q}}_0$ can be bounded elementwise by a pair of iid Gaussian vectors. Since for each of these vectors, the Lipschitz function $\|\theta \cdot\|$ is concentrated about its (very small) expectation, the desired result follows. For the upper block, write $Z_2 = QR$, where $Q \in \mathbb{R}^{\bar\rho m \times \bar\rho m}$ is an orthogonal matrix, and $R \in \mathbb{R}^{\bar\rho m \times (\delta m - k_1)}$ is an upper-triangular matrix with non-negative elements on the diagonal. With probability one (as long as $\mathrm{rank}(Z_2) = \bar\rho m$), $Q$ and $R$ are uniquely determined by $Z_2$. Moreover, $Q$ is a uniform random orthogonal matrix, probabilistically independent of $R$.[13] Since $\hat{\boldsymbol{q}}_0(R_1) = QR\,(R^*R+\mathtt{I})^{-1}\boldsymbol{\vartheta}_{I^c}$ is the product of a uniform random orthogonal matrix and an independent vector $R(R^*R+\mathtt{I})^{-1}\boldsymbol{\vartheta}_{I^c}$, $\frac{\hat{\boldsymbol{q}}_0(R_1)}{\|\hat{\boldsymbol{q}}_0(R_1)\|}$ is uniformly distributed on $\mathbb{S}^{\bar{\rho}m-1}$. With probability $\geq 1-e^{-Cm(1+o(1))}$, $\|\hat{\boldsymbol{q}}_0(R_1)\| = \|Z_2(Z_2^*Z_2+\mathtt{I})\boldsymbol{\vartheta}_{I^c}\| \leq \|Z_2\|\|\pi_{\mathrm{null}(Z_2(Z_2^*Z_2+\mathtt{I})^{-1})^\perp}\boldsymbol{\vartheta}_{I^c}\| \leq 2\nu^2\sqrt{\bar{\rho}}\,(\sqrt{\bar{\rho}}+\sqrt{\delta})\,m^{1/2}$.[14] Introduce an independent random variable $\lambda_1$ distributed as the norm of a $(\bar{\rho}m)$-dimensional iid $\mathcal{N}(0,\sigma^2)$ vector with $\sigma = 4\nu^2(\sqrt{\bar{\rho}}+\sqrt{\delta})$ (i.e., an appropriately scaled $\chi_{\bar{\rho}m}$ rv), and define

$$\boldsymbol{\phi}_1 \doteq \lambda_1 \frac{\hat{\boldsymbol{q}}_0(R_1)}{\|\hat{\boldsymbol{q}}_0(R_1)\|}. \tag{36}$$

Since $\boldsymbol{\phi}_1$ is the product of a uniform random unit vector and an appropriate $\chi$ random variable, its distribution is iid $\mathcal{N}(0,\sigma^2)$. With probability $1-e^{-Cm(1+o(1))}$, $\|\boldsymbol{\phi}_1\| \geq \frac{\sigma}{2}\sqrt{\bar{\rho}m} \geq \|\hat{\boldsymbol{q}}_0(R_1)\|$, so $\boldsymbol{\phi}_1$ dominates $\hat{\boldsymbol{q}}_0(R_1)$ elementwise and $\|\theta\boldsymbol{\phi}_1\| \geq \|\theta\hat{\boldsymbol{q}}_0(R_1)\|$. Applying Lemma 8 of Appendix B, with probability $1-e^{-Cm(1+o(1))}$,

$$\|\theta\boldsymbol{\phi}_1\|_2 \;\leq\; 4\exp\left(-\frac{1}{16\sigma^2}\right)\sqrt{\bar{\rho}m} \;=\; 4\sqrt{\bar{\rho}}\exp\left(-\frac{1}{256\,\nu^4(1+\sqrt{\delta})^2}\right)m^{1/2} \;\leq\; 4\sqrt{\bar{\rho}}\,\exp\left(-\frac{1}{64\nu^2}\right). \tag{37}$$

For the lower ($R_2$) coordinates, write $Z_2^* = [\,Q_1\;\;Q_2\,]\begin{bmatrix} R_1 \\ 0 \end{bmatrix} \doteq QR$ where $R_1 \in \mathbb{R}^{\bar{\rho}m\times\bar{\rho}m}$ is an upper triangular matrix with nonnegative diagonal elements, $Q_1$ is an orthogonal matrix, and $Q_2$ is a random orthobasis for $\mathcal{R}(Q_1)^\perp$ (so that $Q \in \mathbb{R}^{(n-k_1)\times(n-k_1)}$ is an orthogonal matrix). Again from the rotational invariance of the Gaussian distribution, $Q$ is a uniform random orthogonal matrix, independent of $R$, and

$$\hat{\boldsymbol{q}}_0(R_2) = (Z_2^*Z_2+\mathtt{I})^{-1}\boldsymbol{\vartheta}_{I^c} = Q(RR^*+\mathtt{I})^{-1}Q^*\boldsymbol{\vartheta}_{I^c} \doteq Q(RR^*+\mathtt{I})^{-1}\boldsymbol{\gamma}, \tag{38}$$

where $\boldsymbol{\gamma} \doteq Q^*\boldsymbol{\vartheta}_{I^c}$ is an iid $\mathcal{N}(0,\nu^2\rho)$ random vector, *independent of* $Q$. Hence, $\hat{\boldsymbol{q}}_0(R_2)$ is the product of a uniform random orthogonal matrix $Q$, and a probabilistically independent vector $(RR^*+\mathtt{I})^{-1}\boldsymbol{\gamma}$, and its orientation $\frac{\hat{\boldsymbol{q}}_0(R_2)}{\|\hat{\boldsymbol{q}}_0(R_2)\|}$ is a uniform random vector on $\mathbb{S}^{n-k_1-1}$. As above, introduce an independent random variable $\lambda_2$ distributed as the norm of an $(n-k_1)$-dimensional iid $\mathcal{N}(0,4\nu^2\rho)$ random vector, and define

$$\boldsymbol{\phi}_2 = \lambda_2 \frac{\hat{\boldsymbol{q}}_0(R_2)}{\|\hat{\boldsymbol{q}}_0(R_2)\|}. \tag{39}$$

The product of an independent unit vector and (appropriately scaled) $\chi_{n-k_1}$ scalar, $\boldsymbol{\phi}_2$ is distributed as an iid $\mathcal{N}(0,4\nu^2\rho)$ vector. With probability at least $1-e^{-Cm(1+o(1))}$, $\|\boldsymbol{\phi}_2\| \geq \sqrt{2}\nu\sqrt{\rho}\sqrt{n-k_1}$, and

[13] This follows from the rotational invariance of the Gaussian distribution: left multiplication by an independent orthogonal matrix sampled according to the invariant measure yields an independent pair $(Q',R)$ with $Q'R = Z_2' \equiv_d Z_2$.

[14] Here, we have (21) to bound the norm of the projection of $\boldsymbol{\vartheta}$ onto the $(\bar{\rho}m)$-dimensional subspace $\mathrm{null}(Z_2(Z_2^*Z_2+\mathtt{I})^{-1})^\perp$.

$\|\hat{\boldsymbol{q}}_0(R_2)\| \leq \|\boldsymbol{\vartheta}_{I^c}\| \leq \sqrt{2}\nu\sqrt{\rho}\sqrt{n-k_1}$. Therefore, $\boldsymbol{\phi}_2$ dominates $\hat{\boldsymbol{q}}_0(R_2)$ elementwise, and $\|\theta\boldsymbol{\phi}_2\| \geq \|\theta\hat{\boldsymbol{q}}_0(R_2)\|$. By Lemma 8,

$$\|\theta\boldsymbol{\phi}_2\|_2 \leq 4\sqrt{\delta}\exp\left(-\frac{1}{64\nu^2\rho}\right) m^{1/2} \leq 4\sqrt{\delta}\exp\left(-\frac{1}{64\nu^2}\right) m^{1/2} \tag{40}$$

Combining the bounds on $\|\theta\boldsymbol{\phi}_1\|$ and $\|\theta\boldsymbol{\phi}_2\|$ gives the second part of the lemma. ∎

## III. Simulations and Experiments

In this section, we perform simulations verifying the conclusions of Theorem 1, and investigating the effect of various model parameters on the error correction capability of the $\ell^1$-minimization (2). In the simulations below we use the publicly available $\ell^1$-magic package [28], except for one (higher-dimensional) face recognition example, which requires a customized interior point method. Since $\ell^1$-recoverability depends only on the signs and support of $(\boldsymbol{x}_0, \boldsymbol{e}_0)$, in the simulations below we choose $\boldsymbol{x}_0(i) \in \{0,1\}$ and $\boldsymbol{e}_0(i) \in \{-1,0,1\}$. We will judge an output $(\hat{\boldsymbol{x}}, \hat{\boldsymbol{e}})$ to be correct if $\max(\|\hat{\boldsymbol{x}} - \boldsymbol{x}_0\|_\infty, \|\hat{\boldsymbol{e}} - \boldsymbol{e}_0\|_\infty) < 0.01$.

*a) Comparison with alternative approaches:* We first compare the performance of the extended $\ell^1$-minimization

$$\min \ \|\boldsymbol{x}\|_1 + \|\boldsymbol{e}\|_1 \quad \text{subject to} \quad \boldsymbol{y} = A\boldsymbol{x} + \boldsymbol{e}$$

to two alternative approaches. The first is the error correction approach of [14], which multiplies by a full rank matrix $B$ such that $BA = 0$,[15] solves

$$\min \ \|\boldsymbol{e}\|_1 \quad \text{subject to} \quad B\boldsymbol{e} = B\boldsymbol{y},$$

and then subsequently recovers $\boldsymbol{x}$ from the clean system of equations $A\boldsymbol{x} = \boldsymbol{y} - \boldsymbol{e}$. The second is the Regularized Orthogonal Matching Pursuit (ROMP) algorithm [29], a state-of-the-art greedy method for recovering sparse signals.[16] For this algorithm, we use the implementation from `http://math.ucdavis.edu/~dneedell/`.

For this experiment, the ambient dimension is $m = 500$; the parameters of the CAB model are $\nu = 0.05$ and $\delta = 0.25$. We fix the signal support to be $k_1 = 15$, and vary the fraction of errors from 0 to 0.95. For each error fraction, we generate 500 independent problems. Figure 5 plots the fraction of successes

[15]This comparison requires $n \ll m$ although our method is not limited to this case.

[16]For the models considered here, less sophisticated greedy methods such as the standard orthogonal matching pursuit fail even for small error fractions.

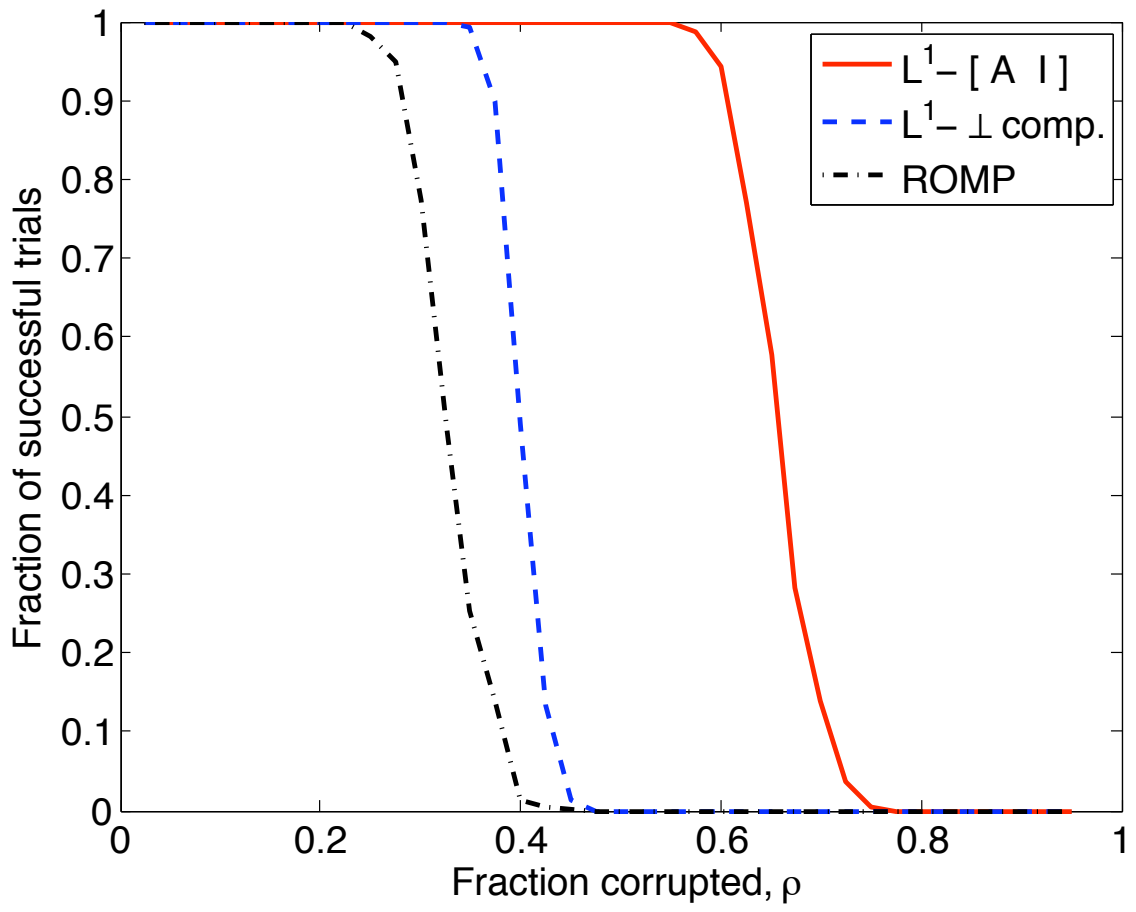

Fig. 5

**Comparison with alternative approaches.** HERE, WE FIX $m = 500$, $\delta = 0.25$, $\nu = 0.05$, AND $k_1 = 15$, AND COMPARE THREE APPROACHES TO RECOVERING THE SPARSE SIGNAL $\boldsymbol{x}_0$ FROM ERROR $\boldsymbol{e}_0$. THE FIRST, DENOTED "$L^1 - [A\ \mathtt{I}]$" SOLVES THE EXTENDED $\ell^1$ MINIMIZATION ADVOCATED IN THIS PAPER. THE SECOND, DENTED "$L^1 - \perp$ COMP" PREMULTIPLIES BY THE ORTHOGONAL COMPLEMENT OF $A$, AND THEN SOLVES AN UNDERDETERMINED SYSTEM OF LINEAR EQUATIONS FOR THE SPARSE ERROR $\boldsymbol{e}$ [14]. THE FINAL APPROACH IS THE GREEDY REGULARIZED ORTHOGONAL MATCHING PURSUIT (ROMP) [29].

for each of the three algorithms, as a function of error density $\rho$. There the extended $\ell^1$-minimization is denoted "$L^1 - [A\ \mathtt{I}]$" (red curve), while the alternative approach of [14] is denoted "$L^1 - \perp$ comp" (blue curve). Whereas both ROMP and the $\ell^1$ approach of [14] break down around $40\%$ corruption, the extended $\ell^1$-minimization continues to succeed with high probability even beyond $60\%$ corruption.

*b) Error correction capacity:* While the previous experiment demonstrates the advantages of the extended $\ell^1$-minimization (2) for the CAB model, Theorem 1 suggests that more is true: As the dimension increases, the fraction of errors that the extended $\ell^1$-minimization can correct should approach one. We generate problem instances with $\delta = 0.25$, $\nu = 0.05$, for varying $m = 100, 200, 400, 800, 1600$. For each problem size, and for each error fraction $\rho = 0.05, 0.1, \dots, 0.95$, we generate 500 random problems, and plot the fraction of correct recoveries in Figure 6. At left, we fix $k_1 = 1$, while at right, $k_1$ grows as $k_1 = m^{1/2}$. In both cases, as $m$ increases, the fraction of errors that can be corrected also increases.

*c) Varying model parameters:* We next investigate the effect of varying $\delta$ (Figure 7 left) and $\nu$ (Figure 7 right). We first fix $m = 400$, $\nu = .3$, and consider different bouquet sizes $n = 100, 200, 300, 400, 500$.

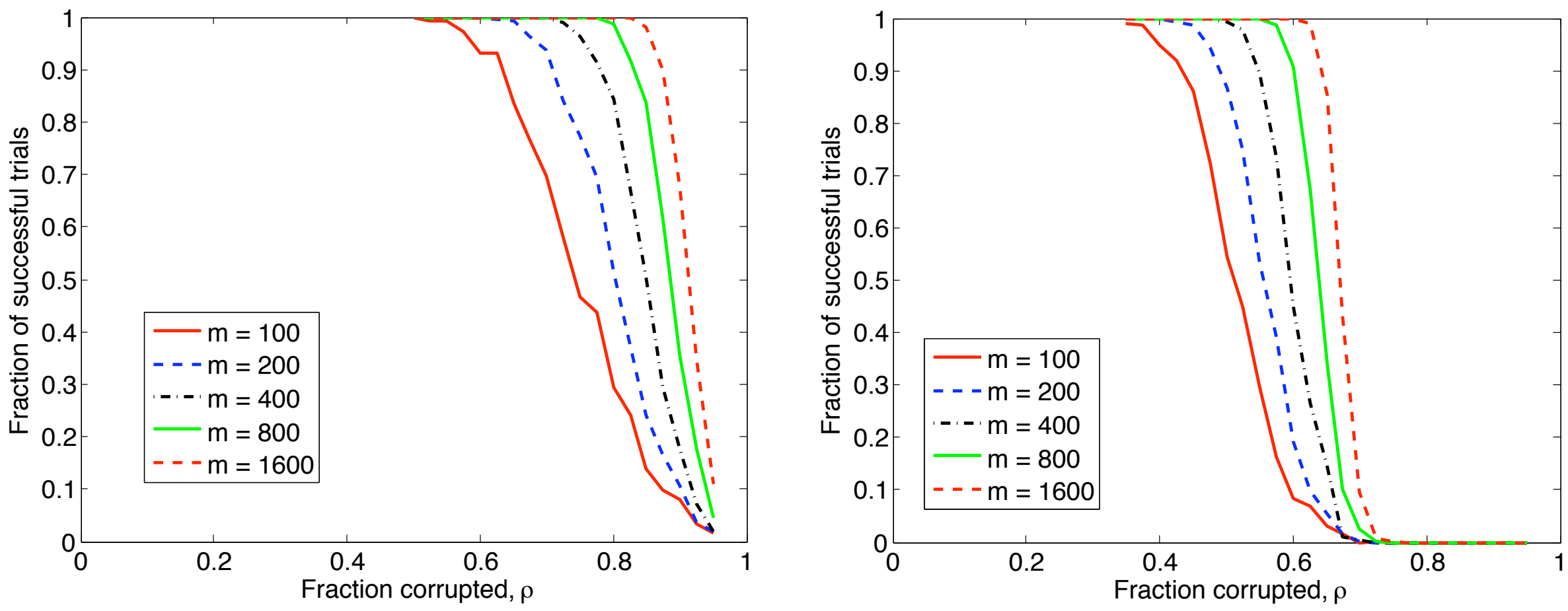

$k_1 = 1$ $\qquad$ $k_1 = m^{1/2}$

Fig. 6

**Error correction in weak proportional growth.** WE FIX $\delta = 0.25$, $\nu = 0.05$, AND PLOT THE FRACTION OF SUCCESSFUL RECOVERIES AS A FUNCTION OF THE ERROR DENSITY $\rho$, FOR EACH $m = 100, 200, 400, 800, 1600$. AT LEFT, $k_1$ IS FIXED AT 1; AT RIGHT, $k_1 = m^{1/2}$. IN BOTH CASES, AS $m$ INCREASES, THE FRACTION OF ERRORS THAT CAN BE CORRECTED APPROACHES 1.

Figure 7 left plots the fraction of correct trials for varying error densities $\rho$, for each of these bouquet sizes. For this fixed $m$, the error correction capability decreases only slightly as $n$ increases.

We next fix $m = 400$, $n = 200$, and consider the effect of varying $\nu$. Figure 7 plots the result for $\nu = .1, .3, .5, .7, .9$. Notice that as $\nu$ decreases (i.e., the bouquet becomes tighter), the error correction capacity increases: for any fixed fraction of successful trials, the fraction of error that can be corrected increases by approximately 15% as $\nu$ decreases from .9 to .5.

*d) Phase transition in total proportional growth:* Theorem 1 does not provide any explicit information about the behavior of $\ell^1$-minimization when the signal support $k_1$ grows proportionally to $m$: $k_1/m \to \rho_1 \in (0, 1)$. Based on intuition from more homogeneous polytopes (especially the work of Donoho and Tanner on Gaussian matrices [24]), we might expect that when $k_1$ also exhibits proportional growth, an asymptotically sharp phase transition between guaranteed recovery and guaranteed failure will occur at some critical error fraction $\rho^* \in (0, 1)$. We investigate this empirically here by again setting $\delta = 0.25$, $\nu = 0.05$, but this time allowing $k_1 = 0.05m$. Figure 8 plots the fraction of correct recovery for varying error fractions $\rho$, as $m$ grows: $m = 100, 200, 400, 800, 1600$. In this proportional growth setting, we see an increasingly sharp phase transition, near $\rho = 0.6$.

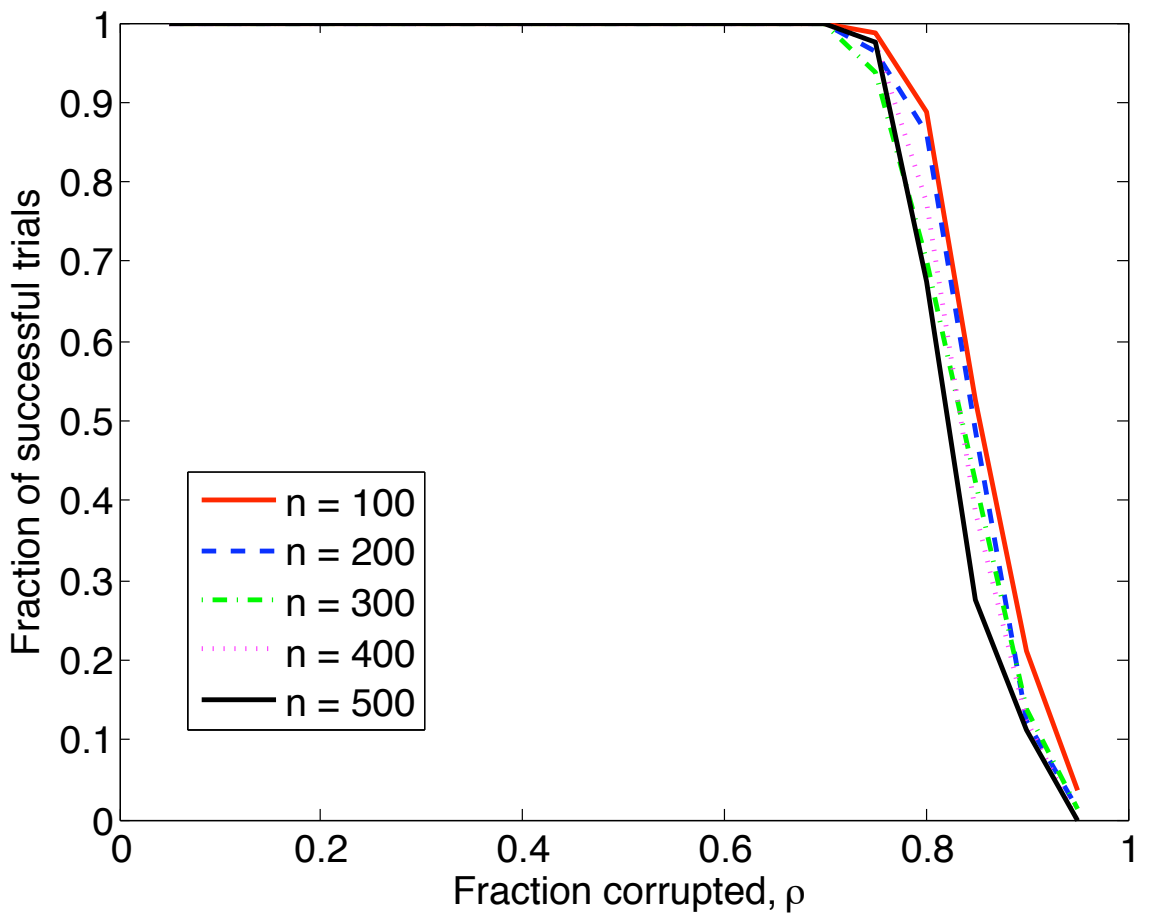

Effect of varying $n$.

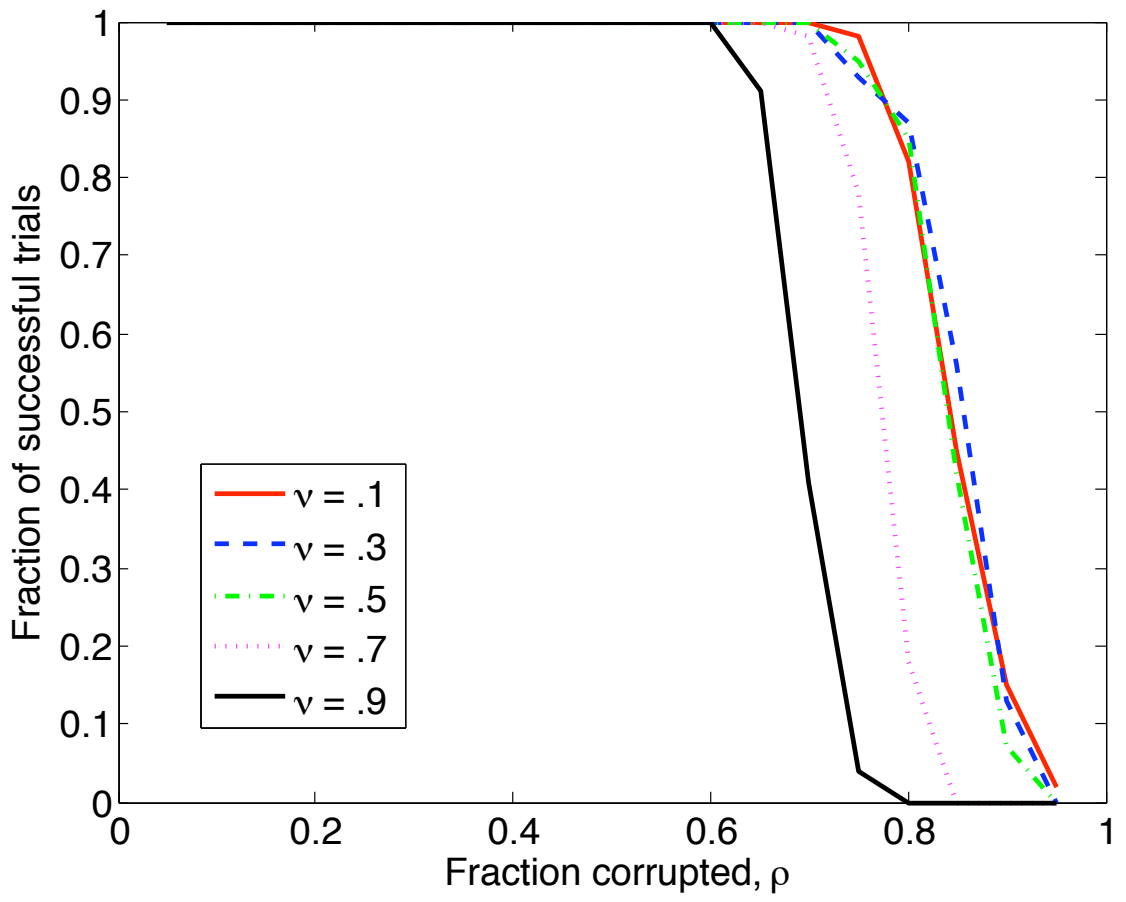

Effect of varying $\nu$.

Fig. 7

**Effect of varying $n$ and $\nu$.** At left, we fix $m = 400$, $\nu = .3$, and consider varying $n = 100, 200, \dots, 500$. For each of these model settings, we plot the fraction of correct recoveries as a function of the fraction of errors. Notice that the error correction capacity decreases only slightly as $n$ increases. At right, we fix $m = 400$, $n = 200$, and vary $\nu$ from .1 to .9. Again, we plot the fraction of correct recoveries for each error fraction. As expected from Theorem 1, as $\nu$ decreases, the error correction capacity of $\ell^1$ increases.

*e) Error correction with real face images:* Finally, we return to the motivating example of face recognition under varying illumination and random corruption. For this experiment, we use the Extended Yale B face database [15], which tests illumination sensitivity of face recognition algorithms. As in [11], we form the matrix $A$ from images in Subsets 1 and 2, which contain mild-to-moderate illumination variations. Each column of the matrix $A$ is a $w \times h$ face image, stacked as a vector in $\mathbb{R}^m$ ($m = w \times h$). Here, the weak proportional growth setting corresponds to the case when the total number of image pixels grows proportionally to the number $n$ of face images. Since the number of images per subject is fixed, this is the same as the total image resolution growing proportionally to the number of subjects. We vary the image resolutions through the range $34 \times 30$, $48 \times 42$, $68 \times 60$, $96 \times 84$.[17] The matrix $A$ is formed from images of $4, 9, 19, 38$ subjects, respectively, corresponding to $\delta \approx 0.09$. Here, $\nu \approx 0.3$. In

[17] Thus, the total dimension $m = 1020, 2016, 4080, 8064$ grows roughly by a factor of 2 from one curve to the next, similar to the simulations above.

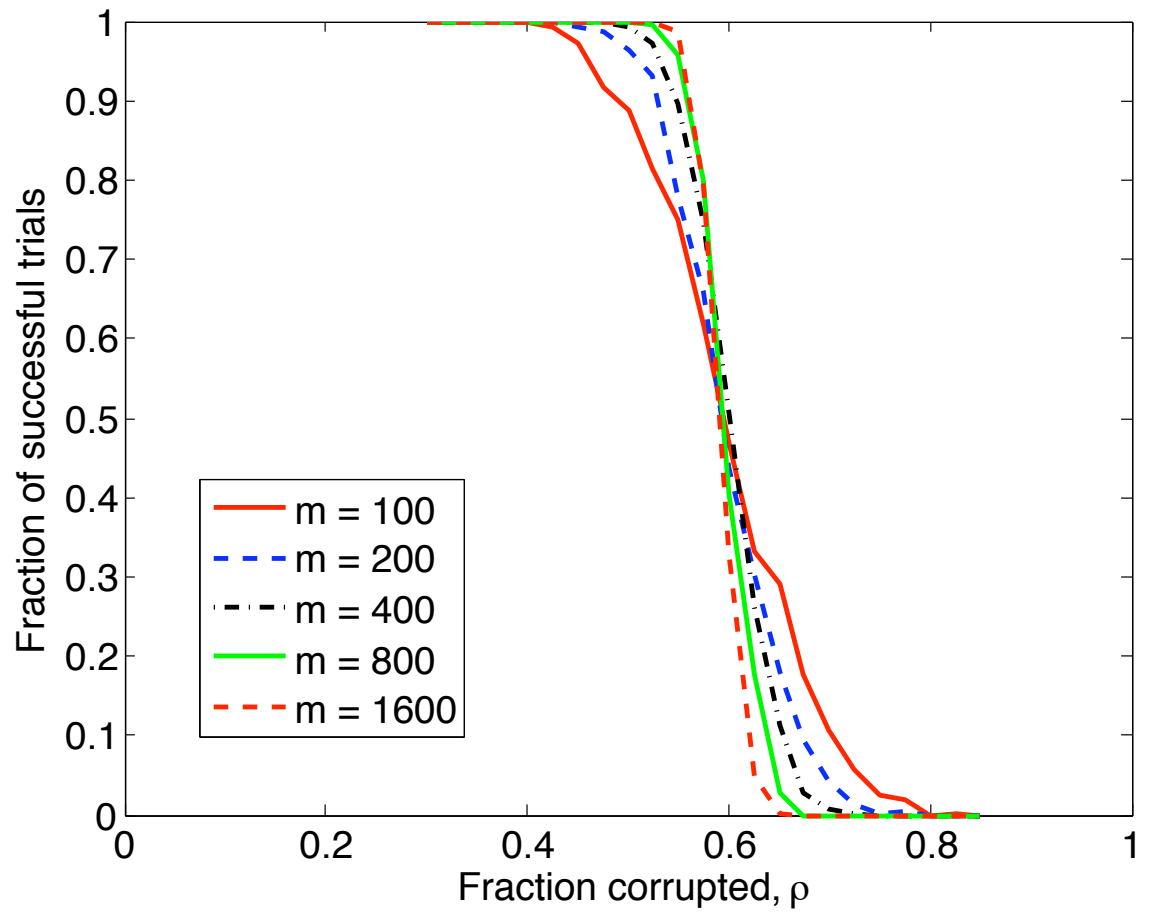

Fig. 8

**Phase transition in total proportional growth.** WHEN THE SIGNAL SUPPORT GROWS IN PROPORTION TO THE DIMENSION ($k_1/m \to \rho_1 \in (0,1)$), WE OBSERVE AN ASYMPTOTICALLY SHARP PHASE TRANSITION IN THE PROBABILITY OF CORRECT RECOVERY, SIMILAR TO THAT INVESTIGATED IN [24]. HERE, FOR $\delta = 0.25$, $\nu = 0.05$, $k_1 = 0.05\,m$, WE INDEED SEE A SHARP PHASE TRANSITION AT $\rho = 0.6$.

face recognition, the sublinear growth of $\|\boldsymbol{x}_0\|_0$ comes from the fact that the observation should ideally be a linear combination of only images of the same subject. Various estimates of the required number of images, $k_1$, appear in the literature, ranging from 5 to 9. Here, we fix $k_1 = 7$, and generate the (clean) test image synthetically as a linear combination of $k_1$ training images from a single subject. The reason for using synthetic linear combinations as opposed to real test images is simply that it allows us to verify whether $\boldsymbol{x}_0$ was correctly recovered; in the real data experiments of the introduction of this paper and of [11], success could only be judged in terms of the recognition rate of the entire classification pipeline.

For each resolution considered, and for each error fraction, we generate 75 trials. Figure 9 (left) plots the fraction of successes as a function of the fraction of corruption. Notice that as predicted by Theorem 1, the fraction of errors that can be corrected again approaches 1 as the data size increases. Figure 9 (right) gives a visual demonstration of the algorithm's capability. In the test images in Figure 9 (right, top), the amount of corruption is chosen to correspond to a 50% probability of success according to the plots in Figure 9 (left). Below each corrupted test image, the "clean" image recovered by our method is shown.

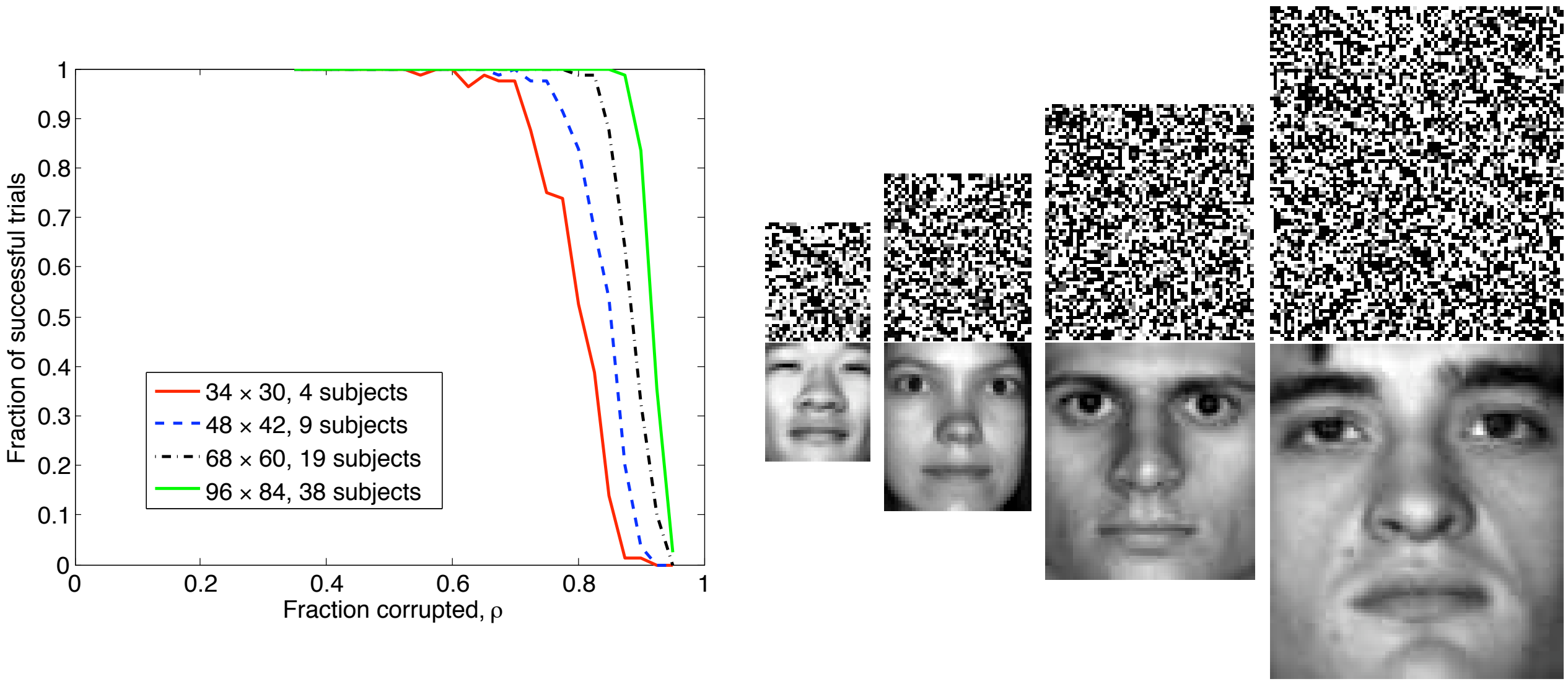

Fig. 9

**Error correction with real face images.** We simulate weak proportional growth in the Extended Yale B face database, with the resolution of the images growing in proportion to the number of subjects. Left: fraction of correct recoveries for varying levels of occlusion. Right: examples of correct recovery for each resolution considered. Top: corrupted test image. The fraction of corruption is chosen so that the probability of correct recovery is 50%. Bottom: clean image, from correctly recovered $\boldsymbol{x}_0$.

## IV. Discussions and Future Work

*a) Compressed sensing for signals with varying sparsity:* In the conventional setting for recovering a sparse signal, one often implicitly assumes that each entry of the signal has an equal probability of being nonzero. As a result, one typically requires that the incoherence (or coherence) of the dictionary is somewhat uniform. In this paper, we saw quite a different example. If we view both $\boldsymbol{x}$ and $\boldsymbol{e}$ as the signal that we want to recover, then the sparsity or density of the combined signal is quite uneven – $\boldsymbol{x}$ is very sparse but $\boldsymbol{e}$ can be very dense. Nevertheless, our result suggests that if the incoherence of the dictionary is adaptive to the distribution of the density – more coherent for the sparse part and less for the dense part, then $\ell^1$-minimization will be able to recover such uneven signals even if bounds based on the even sparsity assumption suggest otherwise. Thus, if one has some prior knowledge about which part of the signal is likely to be more sparse or more dense, one can achieve much better performance with $\ell^1$-minimization by using a dictionary with matching incoherence. More generally, for any given distribution of sparsity, one may ask the question whether there exists an optimal dictionary with matching incoherence such that $\ell^1$-minimization has the highest chance of success.

*b) Stability with respect to noise:* Although in our model, we do not explicitly consider any noise (say $\boldsymbol{y} = A\boldsymbol{x} + \boldsymbol{e} + \boldsymbol{z}$, where $\boldsymbol{z}$ is Gaussian noise), $\ell^1$-minimization is known to be stable under small noise [26]. This is also what we have observed empirically in our simulations and also in experiments with face images: $\ell^1$-minimization for the cross-and-bouquet model is surprisingly stable to measurement or numerical noise. In fact, as the method is able to deal with dense errors regardless of their magnitude, large noisy entries in $\boldsymbol{z}$ will be treated like errors and be absorbed into $\boldsymbol{e}$. However, a more precise characterization of the effect of noise (say Gaussian) on the estimate of the sparse signal $\boldsymbol{x}$ and the error $\boldsymbol{e}$ remains an open problem.

*c) Neighborliness of polytopes:* As we have seen in this paper, a precise characterization of the performance of $\ell^1$-minimization requires us to analyze the geometry of polytopes associated with the specific dictionaries in question. In practice, we often use $\ell^1$-minimization for purposes other than signal reconstruction or error correction. For instance, using machine learning techniques, we can learn from exemplars a dictionary that is optimal for certain tasks such as data classification [13]. The polytope associated with such a dictionary may be very different from those that are normally studied in signal processing or coding theory or error correction, leading to qualitatively different behavior of the $\ell^1$-minimization. Thus, we should expect that in the coming years, many new classes of high-dimensional polytopes with even more interesting properties may arise from other applications and practical problems.

## Acknowledgments

The authors would like to acknowledge helpful conversations with and useful comments from Prof. Robert Fossum (UIUC Math), Prof. Olgica Milenkovic (UIUC ECE), Prof. Sean Meyn (UIUC ECE), and Dr. Gang Hua (Microsoft Live Labs). This work is partially supported by grants NSF CRS-EHS-0509151, NSF CCF-TF-0514955, ONR YIP N00014-05-1-0633, and NSF IIS 07-03756. John Wright is also supported by a Microsoft Fellowship (sponsored by Microsoft Live Labs, Redmond). Finally, Yi Ma would like to thank Microsoft Research Asia, Beijing, China, for its hospitality during his visit there in Summer 2008.

# APPENDIX
# TECHNICAL LEMMAS AND RESULTS

## A. Restricted Isometry for Sparse Vectors

Here, we give a more precise statement of the restricted isometry property of $[\tilde{V}^* \ \ -\tilde{S}\tilde{U}^*]$ used in the proof of Lemma 3. For an arbitrary matrix $M$, we defined $\gamma_k(M) \doteq \inf_{\|\boldsymbol{y}\|_0 \le k,\, \boldsymbol{y} \neq \boldsymbol{0}} \frac{\|M\boldsymbol{y}\|_2}{\|\boldsymbol{y}\|_2}$. We are interested in knowing $\gamma_{cm}([\tilde{V}^* \ \ -\tilde{S}\tilde{U}^*])$, where $\tilde{U}$, $\tilde{S}$, and $\tilde{V}$ come from a (compact) singular value decomposition[18] of $P \doteq A_2^* \pi_{A_1^\perp}$, after dropping the largest singular value. The constants in the following result are less important than the fact that for $c$ sufficiently small, $\gamma_{cm} = \Omega(\nu)$.

*Lemma 5 (Restricted Isometry):* Suppose that $\bar{\rho} < \delta$, $\nu < 1/9$, and $c$ is sufficiently small:

$$c \ \le \ \min\left\{ \frac{\bar{\rho}}{1024}, \frac{\bar{\rho}}{64\,(\,1 + 2C_\mu \bar{\rho}^{-1/2}\,)^2} \right\}, \qquad \bar{\rho} H(c/\bar{\rho}) + \delta H(c/\delta) \ < \ \frac{\bar{\rho}}{128\pi^2}, \tag{41}$$

where $H(\cdot)$ is the base-$e$ binary entropy function. Let $\boldsymbol{u}_1, \boldsymbol{v}_1$ denote the first singular vectors of $P \doteq A_2^* \pi_{A_1^\perp} \in \mathbb{R}^{(\delta m - k_1) \times \bar{\rho} m}$. Then if $\tilde{U}\tilde{S}\tilde{V}^*$ is a compact singular value decomposition of $\pi_{\boldsymbol{u}_1^\perp} P \pi_{\boldsymbol{v}_1^\perp}$,

$$\gamma_{cm}(\,[\tilde{V}^* \ \ -\tilde{S}\tilde{U}^*]\,) \ \ge \ \frac{\nu\sqrt{\bar{\rho}}}{16} \tag{42}$$

on the complement of a bad event of probability $\le e^{-Cm(1+o(1))}$.

*Proof:* Notice that the conditional distribution of $P$ given $A_1$ is Gaussian: $P = Z_2^* \pi_{A_1^\perp} + \boldsymbol{1}\boldsymbol{\mu}_{J^c}^* \pi_{A_1^\perp} \doteq Z_2^* \pi_{A_1^\perp} + \boldsymbol{1}\hat{\boldsymbol{\mu}}^*$. We argue that the second term dominates:

*a) $\boldsymbol{1}\hat{\boldsymbol{\mu}}^*$ determines the leading singular vectors:* Since the columns of $A_1$ are $k_1$ small perturbations of $\boldsymbol{\mu}_{J^c}$, the residual $\|\hat{\boldsymbol{\mu}}\| = \|\pi_{A_1^\perp} \boldsymbol{\mu}_{J^c}\|$ should be small. However, we will see that it is not too small: $\|\pi_{A_1^\perp} \boldsymbol{\mu}_{J^c}\| \ = \ \Omega(k_1^{-1/2})$. Choose an orthonormal basis for $\mathbb{R}^{\bar{\rho} m}$, with first basis vector $\frac{\boldsymbol{\mu}_{J^c}}{\|\boldsymbol{\mu}_{J^c}\|}$. The expression of $A_1$ w.r.t. this basis is $\left[\begin{smallmatrix} \boldsymbol{0} \\ B \end{smallmatrix}\right] + \boldsymbol{e}_1(\boldsymbol{c}^* + \|\boldsymbol{\mu}_{J^c}\| \boldsymbol{1}^*) \ \doteq \ \left[\begin{smallmatrix} \boldsymbol{0} \\ B \end{smallmatrix}\right] + \boldsymbol{e}_1 \boldsymbol{v}^*$, where $B$ and $\boldsymbol{c}$ are iid $\mathcal{N}(0, \nu^2/m)$. So, $\left\| \pi_{A_1} \frac{\boldsymbol{\mu}_{J^c}}{\|\boldsymbol{\mu}_{J^c}\|} \right\|_2^2$ can be written as

$$\boldsymbol{e}_1^* \left( \left[\begin{smallmatrix} \boldsymbol{0} \\ B \end{smallmatrix}\right] + \boldsymbol{e}_1 \boldsymbol{v}^* \right) (\boldsymbol{v}\boldsymbol{v}^* + B^*B)^{-1} \left( \left[\begin{smallmatrix} \boldsymbol{0} & B^* \end{smallmatrix}\right] + \boldsymbol{v}\boldsymbol{e}_1^* \right) \boldsymbol{e}_1 \quad = \quad \frac{\boldsymbol{v}^*(B^*B)^{-1}\boldsymbol{v}}{1 + \boldsymbol{v}^*(B^*B)^{-1}\boldsymbol{v}}.$$

Applying Fact 1 to the $(\bar{\rho}m - 1) \times k_1$ matrix $B$, one can easily show that $P\left[ \|(B^*B)^{-1}\| > \frac{2}{\nu^2 \bar{\rho}} \right] \asymp e^{-Cm}$. By (21) above, the norm of the $k_1$-dimensional $\mathcal{N}(0, \nu^2/m)$ vector $\boldsymbol{c}$ also concentrates: $P\left[ \|\boldsymbol{c}\| > \sqrt{k_1} \right] \asymp e^{-C' m k_1}$. On the complement of these bad events, $\|\boldsymbol{v}\| \le \|\boldsymbol{c}\| + \|\boldsymbol{\mu}_{J^c} \boldsymbol{1}_{k_1}^*\| = (1 + \|\boldsymbol{\mu}_{J^c}\|)\sqrt{k_1} \le 2\sqrt{k_1}$,

[18] With probability one, the matrices $U$ and $V$ are unique upto multiplication of their columns by a common set of signs. The quantity of interest, $\gamma_k$, does not depend on the choice of signs, so there is no ambiguity in writing $\gamma_k([\tilde{V}^* \ \ -\tilde{S}\tilde{U}^*])$.

and $\boldsymbol{v}^*(B^*B)^{-1}\boldsymbol{v} \le \frac{8}{\nu^2\bar{\rho}}k_1$. So,

$$\left\| \frac{\boldsymbol{\mu}_{J^c}}{\|\boldsymbol{\mu}_{J^c}\|} - \pi_{A_1}\frac{\boldsymbol{\mu}_{J^c}}{\|\boldsymbol{\mu}_{J^c}\|} \right\|_2^2 = \frac{1}{1+\boldsymbol{v}^*(B^*B)^{-1}\boldsymbol{v}} \ge \frac{1}{1+\frac{8}{\nu^2\bar{\rho}}k_1}. \tag{43}$$

Lemma 6 below shows that with probability $\ge 1 - e^{-Cm(1+o(1))}$ in the random support of the error $\boldsymbol{e}_0$, $\|\boldsymbol{\mu}_{J^c}\| \ge \bar{\rho}/2$. Together with (43), this implies that $\|\hat{\boldsymbol{\mu}}\| = \|\boldsymbol{\mu}_{J^c} - \pi_{A_1}\boldsymbol{\mu}_{J^c}\|_2 \ge \frac{\bar{\rho}}{2}\sqrt{\frac{1}{1+\frac{8}{\nu^2\bar{\rho}}k_1}}$. On this good event, $\|\mathbf{1}_{\delta m - k_1}\hat{\boldsymbol{\mu}}^*\|_2 \ge C_1\, m^{\eta_0/2}$ for some constant $C_1$ and $m$ sufficiently large. From Fact 1, $\|Z_2\|$ is bounded by some constant $C_2$ with probability at least $1 - e^{-Cm(1+o(1))}$. Treating $Z_2^*\pi_{A_1^\perp}$ as a nuisance perturbation of $\mathbf{1}\hat{\boldsymbol{\mu}}^*$ and applying Wedin's perturbation bound for principal subspaces [30] then gives

$$\begin{aligned}\|\pi_{\boldsymbol{u}_1^\perp} - \pi_{\mathbf{1}^\perp}\| &= \left\| \frac{\boldsymbol{u}_1\boldsymbol{u}_1^*}{\boldsymbol{u}_1^*\boldsymbol{u}_1}\left(\mathtt{I} - \frac{\mathbf{1}\mathbf{1}^*}{\mathbf{1}^*\mathbf{1}} + \frac{\mathbf{1}\mathbf{1}^*}{\mathbf{1}^*\mathbf{1}}\right) - \left(\mathtt{I} - \frac{\boldsymbol{u}_1\boldsymbol{u}_1^*}{\boldsymbol{u}_1^*\boldsymbol{u}_1} + \frac{\boldsymbol{u}_1\boldsymbol{u}_1^*}{\boldsymbol{u}_1^*\boldsymbol{u}_1}\right)\frac{\mathbf{1}\mathbf{1}^*}{\mathbf{1}^*\mathbf{1}} \right\| \\ &\le 2\left\| \frac{\boldsymbol{u}_1\boldsymbol{u}_1^*}{\boldsymbol{u}_1^*\boldsymbol{u}_1}\left(\mathtt{I} - \frac{\mathbf{1}\mathbf{1}^*}{\mathbf{1}^*\mathbf{1}}\right) \right\| \le \frac{2\,\|Z_2^*\pi_{A_1^\perp}\|}{\|\mathbf{1}\hat{\boldsymbol{\mu}}^*\|} \le \frac{2\,C_2}{C_1\, m^{\eta_0/2}}.\end{aligned}$$

Similarly $\|\pi_{\boldsymbol{v}_1^\perp} - \pi_{\hat{\boldsymbol{\mu}}^\perp}\| \le \frac{2\,C_2}{C_1\,m^{\eta_0/2}}$. Write

$$\|\pi_{\boldsymbol{u}_1^\perp}P\pi_{\boldsymbol{v}_1^\perp} - \pi_{\mathbf{1}^\perp}P\pi_{\hat{\boldsymbol{\mu}}^\perp}\| \le \|\pi_{\boldsymbol{u}_1^\perp} - \pi_{\mathbf{1}^\perp}\|\|P\pi_{\boldsymbol{v}_1^\perp}\| + \|\pi_{\mathbf{1}^\perp}P\|\|\pi_{\boldsymbol{v}_1^\perp} - \pi_{\hat{\boldsymbol{\mu}}^\perp}\|.$$

Now, $\|\pi_{\mathbf{1}^\perp}P\| \le \|Z_2\| \le C_2$, and $\|P\pi_{\boldsymbol{v}_1^\perp}\| = \sigma_2(P) \le \sqrt{2}\nu(\sqrt{\bar{\rho}} + \sqrt{\delta})$ simultaneously with probability $\ge 1 - e^{-Cm(1+o(1))}$ (the second bound was established in part (a) of the proof of Lemma 3). Hence, $\exists\, C_3$ such that $P\left[\ \|\pi_{\boldsymbol{u}_1^\perp}P\pi_{\boldsymbol{v}_1^\perp} - \pi_{\mathbf{1}^\perp}P\pi_{\hat{\boldsymbol{\mu}}^\perp}\|_2 > C_3\, m^{-\eta_0/2}\ \right] \asymp e^{-Cm}$. For an arbitrary matrix $W$, let $f(W) \doteq \gamma_{cm}([\pi_{\mathcal{R}(W^*)}\ \ -W^*])$. We are interested in $f(\pi_{\boldsymbol{u}_1^\perp}P\pi_{\boldsymbol{v}_1^\perp})$.[19] Using the fact that singular values of submatrices are 1-Lipschitz and applying Wedin's $\sin\Theta$ theorem [30] to $\pi_{R(W^*)}$, it is not difficult to show that if $\operatorname{rank}(W+\Delta) = \operatorname{rank}(W)$,

$$|\, f(W+\Delta) - f(W)\,| \le \left(\frac{1}{\sigma_{min}(W) - \|\Delta\|} + 1\right)\|\Delta\|, \tag{44}$$

where $\sigma_{min}(W)$ is the smallest nonzero singular value. Applying this bound with $W = \pi_{\boldsymbol{u}_1^\perp}P\pi_{\boldsymbol{v}_1^\perp}$, $\Delta = \pi_{\boldsymbol{u}_1^\perp}P\pi_{\boldsymbol{v}_1^\perp} - \pi_{\mathbf{1}^\perp}P\pi_{\hat{\boldsymbol{\mu}}^\perp}$, and noticing that $\sigma_{min}(\pi_{\boldsymbol{u}_1^\perp}P\pi_{\boldsymbol{v}_1^\perp})$ is bounded below by a positive constant with overwhelming probability, we have that $\left|\ f\left(\pi_{\boldsymbol{u}_1^\perp}P\pi_{\boldsymbol{v}_1^\perp}\right) - f\left(\pi_{\mathbf{1}^\perp}P\pi_{\hat{\boldsymbol{\mu}}^\perp}\right)\ \right| < \frac{\nu\sqrt{\bar{\rho}}}{16}$ with probability at least $1 - e^{-Cm(1+o(1))}$. We henceforth restrict our attention to $f(\pi_{\mathbf{1}^\perp}P\pi_{\hat{\boldsymbol{\mu}}^\perp})$.

*b) Analysis via Gaussian measure concentration:* Let $\Sigma$ denote the subspace $(\mathcal{R}(Z_1) + \mathcal{R}(\boldsymbol{\mu}_{J^c}))^\perp$, and let $V_0$ be some orthonormal basis for this subspace, chosen independently of $Z_2$. From the above reasoning, we can restrict our attention to $\pi_{\mathbf{1}^\perp}P\pi_{\hat{\boldsymbol{\mu}}^\perp} = \pi_{\mathbf{1}^\perp}Z_2^*\pi_\Sigma$. Let $\pi_{\mathbf{1}^\perp}Z_2^*\pi_\Sigma = U'S'V'^*$ be a compact singular value decomposition of this matrix. Then,

$$\gamma_{cm}\left(\begin{bmatrix} V'^* & -S'U'^* \end{bmatrix}\right) = \gamma_{cm}\left(V'^*\begin{bmatrix} \mathtt{I} & \pi_\Sigma Z_2\pi_{1^\perp} \end{bmatrix}\right) = \gamma_{cm}\left(V_0^*\begin{bmatrix} \mathtt{I} & \pi_\Sigma Z_2\pi_{1^\perp} \end{bmatrix}\right).$$

[19] Since left multiplication by an orthogonal matrix does not change $\gamma_{cm}$, $f(\pi_{\boldsymbol{u}_1^\perp}P\pi_{\boldsymbol{v}_1^\perp}) = \gamma_{cm}([\tilde{V}^*\ \ -\tilde{S}\tilde{U}^*])$.

Where the final step follows because $\gamma_{cm}$ is invariant under left multiplication of its argument by an orthogonal matrix. Now, $V_0^*\pi_\Sigma Z_2 = V_0^* Z_2$ is simply distributed as a $(\bar\rho m - k_1 - 1) \times (\delta m - k_1)$ iid $\mathcal{N}(0,\nu^2/m)$ random matrix. Finally, introduce an additional uniformly distributed random orthogonal matrix $Q \in \mathbb{R}^{(\bar\rho m - k_1 - 1)\times(\bar\rho m - k_1 - 1)}$, chosen independently of $Z_2$, and define $\Psi \doteq QV_0^*\pi_\Sigma Z_2$. This is again an iid $\mathcal{N}(0,\nu^2/m)$ matrix. Notice then, that $\gamma_{cm}\left(\left[\begin{array}{cc} V'^* & -S'U'^* \end{array}\right]\right) = \gamma_{cm}\left(\left[\begin{array}{cc} QV_0^* & \Psi\pi_{1^\perp} \end{array}\right]\right)$. From the rotational invariance of the Gaussian distribution, it is easy to show that $\Psi$ and $Q$ are independent random variables. $QV_0^*$ is the transpose of random orthobasis for $\Sigma$; it can be realized by orthogonalizing the projection of a Gaussian matrix onto $\Sigma$. To this end, introduce an iid $\mathcal{N}(0,\nu^2/m)$ matrix $\Phi \in \mathbb{R}^{(\bar\rho m - k_1 - 1)\times\bar\rho m}$ independent of $\Sigma$ and $\Psi$. Then, $\gamma_{cm}\left(\left[\begin{array}{cc} QV_0^* & \Psi\pi_{\mathbf{1}^\perp} \end{array}\right]\right)$ is equal in distribution to $\gamma_{cm}\left(\left[\begin{array}{cc} (\Phi\pi_\Sigma\Phi^*)^{-1/2}\,\Phi\pi_\Sigma & \Psi\pi_{\mathbf{1}^\perp} \end{array}\right]\right)$. Let $\Lambda \doteq (\Phi\pi_\Sigma\Phi^*)^{-1/2}$, and notice that

$$\begin{array}{rcl} \gamma_{cm} & = & \min\limits_{\#L_1\cup L_2 = cm} \sigma_{min}\left(\left[\begin{array}{cc} [\Lambda\Phi\pi_\Sigma]_{\bullet,L_1} & [\Psi\pi_{\mathbf{1}^\perp}]_{\bullet,L_2} \end{array}\right]\right) \quad \geq \\ & & \min\limits_{\#L_1=\#L_2=cm} \min\left\{\sigma_{min}([\Lambda\Phi\pi_\Sigma]_{\bullet,L_1}),\, \sigma_{min}(\pi_{\Sigma'^\perp}[\Psi\pi_{\mathbf{1}^\perp}]_{\bullet,L_2})\right\} \; - \max\limits_{\#L_1=\#L_2=cm}\left\|\pi_{\Sigma'}\left[\Psi\pi_{\mathbf{1}^\perp}\right]_{\bullet,L_2}\right\| \end{array}$$

where $\Sigma'$ denotes the subspace $\mathcal{R}([\Lambda\Phi\pi_\Sigma]_{\bullet,L_1})$.

*c) Bounding* $\sigma_{min}[\Lambda\Phi\pi_\Sigma]_{\bullet,L}$*:* Applying Fact 1 to $\Phi\pi_\Sigma$ gives that $P\left[\,\|\Phi\pi_\Sigma\|_2 \geq 3\nu\sqrt{\bar\rho}\,\right] \asymp e^{-\bar\rho m/2}$. On the complement of this bad event, $\sigma_{min}(\Lambda) \geq \frac{1}{3\nu\sqrt{\bar\rho}}$. Write

$$\begin{array}{rcl} [\Phi\pi_\Sigma]_{\bullet,L} & = & \Phi_{\bullet,L} - [\Phi\pi_{\Sigma^\perp}]_{\bullet,L} \;=\; \Phi_{\bullet,L}(\mathtt{I} - [\pi_{\Sigma^\perp}]_{L,L}) - \Phi_{\bullet,L^c}[\pi_\Sigma]_{L^c,L} \\ \Longrightarrow \quad \sigma_{min}([\Phi\pi_\Sigma]_{\bullet,L}) & \geq & \sigma_{min}(\Phi_{\bullet,L})(1 - \|[\pi_{\Sigma^\perp}]_{L,L}\|) - \|\pi_{\Phi_{\bullet,L}}\Phi_{\bullet,L^c}\pi_{[\pi_\Sigma]_{L^c,L}}\|. \end{array}$$

Straightforward application of Fact 1 shows that $P\left[\sigma_{min}(\Phi_{\bullet,L}) \leq \frac{\nu\sqrt{\bar\rho}}{2} - \nu\sqrt{c}\right] \asymp e^{-\bar\rho m/8}$, while for any[20] $\varepsilon_1 > 0$, $P\left[\|\pi_{\Phi_{\bullet,L}}\Phi_{\bullet,L^c}\pi_{[\pi_\Sigma]_{L^c,L}}\| \geq 2\nu\sqrt{c} + \nu\sqrt{\bar\rho\varepsilon_1}\right] \asymp e^{-\bar\rho\varepsilon_1 m/2}$. Finally, consider the matrix $\Upsilon \doteq \left[Z_1 \quad \nu\sqrt{\bar\rho}\frac{\boldsymbol{\mu}_{J^c}}{\|\boldsymbol{\mu}_{J^c}\|}\right] \in \mathbb{R}^{\bar\rho m\times(k_1+1)}$. We are interested in $\|[\pi_{\Sigma^\perp}]_{L,L}\| = \left\|\Upsilon_{L,\bullet}(\Upsilon^*\Upsilon)^{-1}\Upsilon^*_{\bullet,L}\right\| \leq \frac{\|\Upsilon_{L,\bullet}\|^2}{\sigma^2_{min}(\Upsilon)}$. It is not difficult to show[21] that w.p. $\geq 1 - e^{-\frac{\bar\rho m}{8}(1-\varepsilon+o(1))}$, $\sigma_{min}(\Upsilon) \geq \frac{\nu\sqrt{\bar\rho}}{2}$. Meanwhile for any $\varepsilon_2 > 0$, $P\left[\,\|[Z_1]_{L,\bullet}\| \geq \nu\sqrt{c} + \nu\sqrt{\bar\rho\varepsilon_2}\right] \asymp e^{-\bar\rho\varepsilon_2 m/2}$. On the complement of this bad event (and invoking Lemma 6)

$$\|\Upsilon_{L,\bullet}\| \;\leq\; \|[Z_1]_{L,\bullet}\| + \left\|\nu\sqrt{\bar\rho}\frac{\boldsymbol{\mu}_{J^c}(L)}{\|\boldsymbol{\mu}_{J^c}\|}\right\| \;\leq\; \nu\sqrt{c} + \nu\sqrt{\bar\rho\varepsilon_2} + 2\nu C_\mu\sqrt{\frac{c}{\bar\rho}} \;=\; \nu\left(\sqrt{\bar\rho\varepsilon_2} + \sqrt{c}\left(1 + \frac{2C_\mu}{\sqrt{\bar\rho}}\right)\right).$$

[20]Since $\Phi_{\bullet,L^c}$ is independent of $\Phi_{\bullet,L}$ and $\Sigma$, the norm of $\pi_{\Phi_{\bullet,L}}\Phi_{\bullet,L^c}\pi_{[\pi_\Sigma]_{L^c,L}}$ is simply distributed as the norm of a $cm\times cm$ iid $\mathcal{N}(0,\nu^2/m)$ matrix. By Fact 1, $P\left[\|\pi_{\Phi_{\bullet,L}}\Phi_{\bullet,L^c}\pi_{[\pi_\Sigma]_{L^c,L}}\| \geq 2\nu\sqrt{c} + t\nu\sqrt{c}\right] \leq e^{-(t-o(1))^2 cm/2}$. Set $t = \sqrt{\frac{\bar\rho\varepsilon_1}{c}}$.

[21]Write $\sigma_{min}(\Upsilon) \geq \sigma_{min}\left(\left[\begin{array}{cc} \pi_{\boldsymbol{\mu}^\perp_{J^c}}Z_1 & \nu\sqrt{\bar\rho}\frac{\boldsymbol{\mu}_{J^c}}{\|\boldsymbol{\mu}_{J^c}\|} \end{array}\right]\right) - \|\pi_{\boldsymbol{\mu}_{J^c}}Z_1\| \geq \min\left(\sigma_{min}(\pi_{\boldsymbol{\mu}^\perp_{J^c}}Z_1), \nu\sqrt{\bar\rho}\right) - \|\pi_{\boldsymbol{\mu}_{J^c}}Z_1\|$, apply Fact 1 to the singular value and standard tail bounds to the $k_1$-dimensional $\mathcal{N}(0,\nu^2/m)$ vector $\frac{\boldsymbol{\mu}^*_{J^c}}{\|\boldsymbol{\mu}_{J^c}\|}Z_1$.

By the assumptions of the lemma, $\sqrt{c}\,(1+\frac{2C_\mu}{\sqrt{\bar\rho}}) \le \sqrt{\bar\rho}/8$, and $\|[\pi_{\Sigma^\perp}]_{L,L}\| \le \frac{\|\Upsilon_{L,\bullet}\|^2}{\sigma^2_{min}(\Upsilon)} \le 4\,(\sqrt{\varepsilon_2}+1/8)^2$. Setting $\varepsilon_1=\varepsilon_2=\frac{1}{64}$, $\|[\pi_{\Sigma^\perp}]_{L,L}\| \le 1/4$, $\|\pi_{\Phi_{\bullet,L}}\Phi_{\bullet,L^c}\pi_{[\pi_\Sigma]_{L^c,L}}\| \le 2\nu\sqrt{c}+\nu\sqrt{\bar\rho}/8$, and so

$$\sigma_{min}\left([\Phi\pi_\Sigma]_{\bullet,L}\right) \;\ge\; \left(\frac{\nu\sqrt{\bar\rho}}{2}-\nu\sqrt{c}\right)\left(1-\frac{1}{4}\right)-\left(2\,\nu\sqrt{c}+\frac{\nu\sqrt{\bar\rho}}{8}\right) \;=\; \frac{\nu\sqrt{\bar\rho}}{4}-\frac{11\nu\sqrt{c}}{4}, \tag{45}$$

and $\sigma_{min}\left([\Lambda\Phi\pi_\Sigma]_{\bullet,L}\right) \ge \frac{1}{12}-\frac{11}{12}\sqrt{\frac{c}{\bar\rho}} > \frac{1}{24}$ on the complement of a bad event of probability $e^{-\frac{\bar\rho m}{128}(1+o(1))}$. The number of subsets $L$ of size $cm$ is $e^{\bar\rho m H(c/\bar\rho)(1+o(1))}$. The probability any $L$ is bad is bounded by $e^{\bar\rho m\left(H(c/\bar\rho)-\frac{1}{128}\right)(1+o(1))}$, which falls off exponentially when $H(c/\bar\rho) < 1/128$. This is guaranteed for $c/\bar\rho < 1/1024$.

*d) Bounding $\sigma_{min}\,(\pi_{\Sigma'^\perp}[\Psi\pi_{\mathbf{1}^\perp}]_{\bullet,L})$:* Recall that $\Sigma'$ denotes the $cm$-dimensional range of $[\Lambda\Phi\pi_\Sigma]_{\bullet,L_1}$. Choose any orthonormal basis for the $[(\bar\rho-c)m-k_1-1]$-dimensional subspace $\Sigma'^\perp$. The expression of the columns of $\pi_{\Sigma'^\perp}\Psi$ with respect to this basis is a $((\bar\rho-c)m-k_1-1)\times(\delta m-k_1)$ matrix $\tilde\Psi$ with entries $\mathcal{N}(0,\nu^2/m)$. Split $\tilde\Psi\pi_{\mathbf{1}^\perp}$ as

$$[\tilde\Psi\pi_{\mathbf{1}^\perp}]_{\bullet,L} \;=\; \tilde\Psi_{\bullet,L}-\frac{1}{\delta m-k_1}\tilde\Psi_{\bullet,L^c}\mathbf{1}_{L^c}\mathbf{1}_L^*-\frac{1}{\delta m-k_1}\tilde\Psi_{\bullet,L}\mathbf{1}_L\mathbf{1}_L^*.$$

Using the independence of $\frac{1}{\delta m-k_1}\tilde\Psi_{\bullet,L^c}\mathbf{1}$ and $\tilde\Psi_{\bullet,L}$ and applying Fact 1, it is not difficult to show[22] that

$$P\left[\sigma_{min}\left(\tilde\Psi_{\bullet,L}-\frac{1}{\delta m-k_1}\tilde\Psi_{\bullet,L^c}\mathbf{1}\mathbf{1}^*\right) \le \frac{\nu\sqrt{\bar\rho-c}}{2}-\nu\sqrt{c}\right] \le e^{-\frac{(\bar\rho-c)m}{8}(1+o(1))}. \tag{46}$$

For the other term, $\left\|\frac{1}{\delta m-k_1}\tilde\Psi_{\bullet,L}\mathbf{1}\mathbf{1}^*\right\| \;\le\; \|\tilde\Psi_{\bullet,L}\|\frac{c}{\delta}(1+o(1))$. From Fact 1, $P\left[\|\tilde\Psi_{\bullet,L}\| \ge 3\nu\sqrt{\bar\rho}\right] \le e^{-\frac{(\bar\rho-c)m}{2}(1+o(1))}$. On the complement of this event, $\left\|\frac{1}{\delta m-k_1}\tilde\Psi_{\bullet,L}\mathbf{1}\mathbf{1}^*\right\| \le \frac{3\nu c\sqrt{\bar\rho}}{\delta}(1+o(1)) \le 4\nu\sqrt{c}\sqrt{\frac{c}{\delta}}$ eventually. Since $\sqrt{\frac{c}{\delta}} < \sqrt{\frac{c}{\bar\rho}} \le \frac{1}{32}$, $\left\|\frac{1}{\delta m-k_1}\tilde\Psi_{\bullet,L}\mathbf{1}\mathbf{1}^*\right\| \le \frac{\nu\sqrt{c}}{8}$. All together, with probability at least $1-e^{-(\bar\rho-c)m/8(1+o(1))}$,

$$\sigma_{min}([\tilde\Psi\pi_{\mathbf{1}^\perp}]_{\bullet,L}) \;\ge\; \sigma_{min}\left(\tilde\Psi_{\bullet,L}-\frac{1}{m}\tilde\Psi_{\bullet,L^c}\mathbf{1}\mathbf{1}^*\right)-\left\|\frac{\tilde\Psi_{\bullet,L}\mathbf{1}\mathbf{1}^*}{\delta m-k_1}\right\| \;\ge\; \frac{7\nu\sqrt{\bar\rho}}{16}-\frac{9}{8}\nu\sqrt{c} \;\ge\; \frac{3}{8}\nu\sqrt{\bar\rho}$$

There are $\asymp e^{\bar\rho m H(c/\bar\rho)}$ subsets $L_1$ of size $cm$ and $\asymp e^{\delta m H(c/\delta)}$ subsets $L_2$ of size $cm$. The total number of choices of $L_1, L_2$ is asymptotic to $e^{\left(\bar\rho H\left(\frac{c}{\bar\rho}\right)+\delta H\left(\frac{c}{\delta}\right)\right)m}$, and the probability that any pair is bad is bounded by a function asymptotic to $\exp\Big(\big(\bar\rho H(c/\bar\rho)+\delta H(c/\delta)-\frac{\bar\rho-c}{8}\big)\,m\,(1+o(1))\Big)$. Under the assumptions of the lemma, the exponent is negative.

[22]Translation does not substantially affect the bound on $\sigma_{min}$ in Fact 1: for an $m\times n$ iid $\mathcal{N}(0,1/m)$ matrix $M$ and an independent translation $\boldsymbol{x}$, $\sigma_{min}(M+\boldsymbol{x}\mathbf{1}^*) \ge \sigma_{min}(\pi_{\boldsymbol{x}^\perp}M)$, which obeys the same concentration result, now applied to an $(m-1)\times n$ matrix. Appropriate rescaling of the $((\bar\rho-c)m-k_1-1)\times cm$ $\mathcal{N}(0,\nu^2/m)$ matrix $\tilde\Psi_{\bullet,L}$ yields the desired expression.

*e) Bounding the cross-coherence* $\left\|\pi_{\Sigma''}\left[\Psi\pi_{\mathbf{1}^\perp}\right]_{\bullet,L_2}\right\|$*:* Let $\Sigma''$ denote the subspace $\mathcal{R}([\Lambda\Phi\pi_\Sigma]_{\bullet,L_1})$. Notice that $\Sigma''$ and $\Psi$ are probabilistically independent. Now,

$$\left\|\pi_{\Sigma''}\left[\Psi\pi_{\mathbf{1}^\perp}\right]_{\bullet,L_2}\right\| \;\leq\; \|\pi_{\Sigma''}\Psi_{\bullet,L_2}\| + \left\|\pi_{\Sigma''}\Psi\frac{\mathbf{1}_{\delta m-k_1}\mathbf{1}_{cm}^*}{\delta m-k_1}\right\| \;\leq\; \|\pi_{\Sigma''}\Psi_{\bullet,L_2}\| \,+\, \left\|\frac{\pi_{\Sigma''}\Psi\mathbf{1}}{\sqrt{\delta m-k_1}}\right\|$$

eventually, since $\frac{\|\mathbf{1}_{cm}\|}{\sqrt{\delta m-k_1}} = \sqrt{\frac{cm}{\delta m-k_1}} < 1$ eventually. Now, $\|\pi_{\Sigma''}\Psi_{\bullet,L_2}\|$ is distributed as the norm of a $cm\times cm$ iid $\mathcal{N}(0,\nu^2/m)$ matrix, and so for any $\varepsilon_1>0$,

$$P\left[\,\|\pi_{\Sigma''}\Psi_{\bullet,L_2}\| \,\geq\, 2\nu\sqrt{c} \,+\, \varepsilon_1\nu\sqrt{\bar\rho}\,\right] \;\asymp\; e^{-\varepsilon_1^2\bar\rho m/2}. \tag{47}$$

Similarly, $\frac{1}{\sqrt{\delta m-k_1}}\pi_{\Sigma''}\Psi\mathbf{1}$ is has the same norm as a $cm$-dimensional iid $\mathcal{N}(0,\nu^2/m)$ vector, so

$$P\left[\left\|\frac{1}{\sqrt{\delta m-k_1}}\pi_{\Sigma''}\Psi\mathbf{1}\right\| \geq \nu\sqrt{c}+\varepsilon_2\nu\sqrt{\bar\rho}\right] \;\leq\; e^{-2\varepsilon_2^2\bar\rho m/\pi^2}. \tag{48}$$

On the complement of these two bad events, $\left\|\pi_{\Sigma''}\left[\Psi\pi_{1^\perp}\right]_{\bullet,L_2}\right\| \;\leq\; (\varepsilon_1+\varepsilon_2)\,\nu\sqrt{\bar\rho} + 3\,\nu\sqrt{c}$. Set $\varepsilon_1 = \varepsilon_2 = 1/16$. Then w.p. $\geq 1 - e^{-\frac{\bar\rho m}{128\pi^2}(1+o(1))}$, $\left\|\pi_{\Sigma''}\left[\Psi\pi_{1^\perp}\right]_{\bullet,L_2}\right\| \leq \frac{\nu\sqrt{\bar\rho}}{8} + 3\nu\sqrt{c} \leq \frac{\nu\sqrt{\bar\rho}}{4}$. We again union bound over $L_1, L_2$. The number of such pairs is asymptotic to $e^{\left(\bar\rho H\left(\frac{c}{\bar\rho}\right)+\delta H\left(\frac{c}{\delta}\right)\right)m}$, and the probability of some bad pair is bounded by a function asymptotic to $\exp\left(\left(\bar\rho H\left(\frac{c}{\bar\rho}\right) + \delta H\left(\frac{c}{\delta}\right) - \frac{\bar\rho}{128\,\pi^2}\right)m\right)$. Under the hypotheses of the lemma, the coefficient of this exponent is negative.

*f) Pulling the bounds together:* For $\nu < 1/9$, $\frac{3\nu\sqrt{\bar\rho}}{8} < \frac{1}{24} \leq \min_{L_1}\sigma_{min}([\Lambda\Phi\pi_\Sigma]_{\bullet,L_1})$, and so this quantity lower bounds $\min_{L_1,L_2}\min\Big\{\sigma_{min}([\Lambda\Phi\pi_\Sigma]_{\bullet,L_1}),\,\sigma_{min}(\pi_{\Sigma'^\perp}[\Psi\pi_{\mathbf{1}^\perp}]_{\bullet,L_2})\Big\}$. So, w.p. $\geq 1 - e^{-Cm(1+o(1))}$, $\gamma_{cm}\big(\big[\,(\Phi\pi_\Sigma\Phi^*)^{-1/2}\Phi\pi_\Sigma\;\;\Psi\pi_{\mathbf{1}^\perp}\,\big]\big) \;\geq\; \frac{3}{8}\nu\sqrt{\bar\rho} - \frac{1}{4}\nu\sqrt{\bar\rho} \;=\; \frac{\nu\sqrt{\bar\rho}}{8}$. Since

$$\left|\gamma_{cm}\left(\left[\,\tilde V^* \;-\tilde S\tilde U^*\,\right]\right) - \gamma_{cm}\left(\left[\,(\Phi\pi_\Sigma\Phi^*)^{-1/2}\Phi\pi_\Sigma\;\;\Psi\pi_{\mathbf{1}^\perp}\,\right]\right)\right| \;\leq\; \frac{\nu\sqrt{\bar\rho}}{16},$$

the desired bound follows. ■

*Lemma 6:* Let $J^c$ be chosen uniformly at random from $\binom{[m]}{\bar\rho m}$, and let $\boldsymbol{\mu}\in\mathbb{R}^m$ with $\|\boldsymbol{\mu}\|_2 = 1$ and $\|\boldsymbol{\mu}\|_\infty \leq C_\mu m^{-1/2}$. Then $\|\boldsymbol{\mu}_{J^c}\|_2 \geq \bar\rho/2$ on the complement of a bad event of probability $\leq e^{-Cm(1+o(1))}$.

*Proof:* Form the subset $J^c$ by choosing $\bar\rho m$ indices $j_1\ldots j_{\bar\rho m}$, with $j_i$ chosen uniformly at random from $[m]\setminus\{j_1\ldots j_{i-1}\}$. Let $Y_0, Y_1, \ldots Y_{\bar\rho m}$ denote the Doob process associated with $\|\boldsymbol{\mu}_{J^c}\|_2^2$: $Y_0 \doteq \mathbb{E}\left[\,\|\boldsymbol{\mu}_{J^c}\|_2^2\,\right] = \bar\rho$ and $Y_k \doteq \mathbb{E}\left[\|\boldsymbol{\mu}_{J^c}\|_2^2 \mid j_1\ldots j_k\right]$. Then, letting $X_k \doteq \sum_{i=1}^k \boldsymbol{\mu}_{j_i}^2$, $Y_k = X_k + \frac{1-X_k}{m-k}(\bar\rho m - k) = \frac{\rho m X_k+1}{m-k}$, and

$$|Y_{k+1}-Y_k| \;=\; \left|\frac{\rho m(X_k+\boldsymbol{\mu}_{j_{k+1}}^2)+1}{m-k-1} - \frac{\rho m X_k+1}{m-k}\right| \;\leq\; \frac{\rho m X_k + \rho^2m^2\boldsymbol{\mu}_{j_{k+1}}^2+1}{\rho^2 m^2} \;\leq\; \frac{1}{\rho m}+\frac{C_\mu}{m}+\frac{1}{\rho^2m^2}.$$

The above is $\leq C'm^{-1}$ for appropriate constant $C'$. By Azuma's inequality (Theorem 7.2.1 of [31]),

$$P\left[\,|Y_{\bar\rho m}-\bar\rho| \,\geq\, t\,\right] \;\leq\; 2\exp\left(-\frac{t^2}{2\bar\rho m(C'/m)^2}\right) \;\asymp\; \exp(-Cm). \tag{49}$$

■

## B. Technical Lemmas for Initial Separating Hyperplane

This section contains two results used above for controlling the initial separator $\boldsymbol{q}_0$. We first justify the assertion that $\left[\begin{smallmatrix} Z_1 & Z_2 \\ 0 & \mathrm{I} \end{smallmatrix}\right](G^*G)^{-1}Z_{J,\bullet}^*\boldsymbol{\sigma}$ is the only term that contributes $O(m^{1/2})$ to $\|\boldsymbol{q}_0\|$, and then close with a measure concentration result for $\|\theta\cdot\|$, also used in the proof of Lemma 4.

*Lemma 7 (Lower order terms in $\boldsymbol{q}_0$):* Suppose that $\bar{\rho}<\delta$ and $\nu<\frac{1}{8(\sqrt{\delta}+1)}$. There exist constants (wrt $m$) $C_G$ and $C_q$ such that

$$\|(G^*G)^{-1}\| \;\le\; C_G \qquad \text{and} \qquad \left\|\boldsymbol{q}_0-\left[\begin{smallmatrix} Z_1 & Z_2 \\ 0 & \mathrm{I} \end{smallmatrix}\right](G^*G)^{-1}Z_{J,\bullet}^*\boldsymbol{\sigma}\right\| \;\le\; C_q\, m^{1/2-\eta_0/4} \tag{50}$$

simultaneously on the complement of a bad event of probability $\le e^{-Cm^{1-\eta_0/2}(1+o(1))}$.

*Proof:* Write $Q=\left[\begin{smallmatrix} Z_1^*Z_1 & Z_1^*Z_2 \\ Z_2^*Z_1 & Z_2^*Z_2+\mathrm{I} \end{smallmatrix}\right]\in\mathbb{R}^{n\times n}$, and $\boldsymbol{\zeta}=Z_{J^c,\bullet}^*\boldsymbol{\mu}_{J^c}\in\mathbb{R}^n$. Then $G^*G=Q+\boldsymbol{\zeta}\mathbf{1}^*+\mathbf{1}\boldsymbol{\zeta}^*+\alpha\mathbf{1}\mathbf{1}^*$, where $\alpha=\boldsymbol{\mu}_{J^c}^*\boldsymbol{\mu}_{J^c}$. So,

$$(G^*G)^{-1}=Q^{-1}-Q^{-1}\begin{bmatrix}\mathbf{1} & \boldsymbol{\zeta}\end{bmatrix}\begin{bmatrix}\mathbf{1}^*Q^{-1}\mathbf{1} & \mathbf{1}^*Q^{-1}\boldsymbol{\zeta}+1 \\ \mathbf{1}^*Q^{-1}\boldsymbol{\zeta}+1 & \boldsymbol{\zeta}^*Q^{-1}\boldsymbol{\zeta}-\alpha\end{bmatrix}^{-1}\begin{bmatrix}\mathbf{1}^* \\ \boldsymbol{\zeta}^*\end{bmatrix}Q^{-1}. \tag{51}$$

Set $b\doteq\mathbf{1}^*Q^{-1}\mathbf{1}$, $c\doteq\mathbf{1}^*Q^{-1}\boldsymbol{\zeta}$, $d\doteq\boldsymbol{\zeta}^*Q^{-1}\boldsymbol{\zeta}$, and write $(G^*G)^{-1}=Q^{-1}-Q^{-1/2}M\Xi M^*Q^{-1/2}$ with

$$M=\begin{bmatrix}\frac{Q^{-1/2}\mathbf{1}}{\|Q^{-1/2}\mathbf{1}\|_2} & \frac{Q^{-1/2}\boldsymbol{\zeta}}{\|Q^{-1/2}\boldsymbol{\zeta}\|_2}\end{bmatrix} \qquad \text{and} \qquad \Xi=\frac{\begin{bmatrix} b\,(\alpha-d) & -\sqrt{bd}(c+1) \\ -\sqrt{bd}(c+1) & bd\end{bmatrix}}{b\,(\alpha-d)+(c+1)^2}. \tag{52}$$

We next bound the quadratic terms $b$, $c$, and $d$. Applying Fact 1 to the $\delta m\times\bar{\rho}m$ iid $\mathcal{N}(0,\nu^2/m)$ matrix $Z_{J^c,\bullet}=[Z_1\ Z_2]$ gives that $\|Z_{J^c,\bullet}\|_2\le\sqrt{2}\nu\left(\sqrt{\delta}+\sqrt{\bar{\rho}}\right)$ w.p. $\ge 1-e^{-Cm(1+o(1))}$. On the complement of that bad event,

$$b \;=\; \mathbf{1}^*Q^{-1}\mathbf{1} \;\ge\; \frac{\|\mathbf{1}\|_2^2}{\|Q\|} \;\ge\; \frac{\delta m}{1+\|Z_{J^c,\bullet}\|^2} \;\ge\; \frac{\delta m}{1+2\,\nu^2\,(\sqrt{\delta}+\sqrt{\bar{\rho}})^2} \;\doteq\; C_b\, m. \tag{53}$$

Similarly, $b\le\delta m/\sigma_{min}(Q)$. It is not difficult to show[23] that for any block matrix $M=\left[\begin{smallmatrix} A & B \\ 0 & \mathrm{I} \end{smallmatrix}\right]$ with $\sigma_{min}(A)<1$,

$$\sigma_{min}^2(M) \;\ge\; \sigma_{min}^2(A)-\frac{\|A\|^2\,\|B\|^2}{1-\sigma_{min}^2(A)}.$$

By Fact 1, on the complement of an event of probability $\asymp e^{-Cm}$,

$$\sigma_{min}^2(Z_1)\ge\frac{\nu^2\bar{\rho}}{2}, \qquad \|Z_1\|^2\le\|Z_1\|^2\le 2\nu^2\bar{\rho}, \qquad \|Z_2\|^2\le\|Z_2\|^2\le 2\,\nu^2\left(\sqrt{\delta}+\sqrt{\bar{\rho}}\right)^2.$$

On the good event above, for $\nu<\frac{1}{\sqrt{2}}$, $\sigma_{min}^2(Z_1)\le\|Z_1\|^2<1$. Plugging in, $\sigma_{min}(Q)=\sigma_{min}^2\left(\left[\begin{smallmatrix} Z_1 & Z_2 \\ 0 & \mathrm{I} \end{smallmatrix}\right]\right)\ge\sigma_{min}^2(Z_1)-\frac{\|Z_1\|^2\|Z_2\|^2}{1-\sigma_{min}^2(Z_1)}\ge\frac{\nu^2\bar{\rho}}{2}-\frac{4\,\nu^4\bar{\rho}\,(\sqrt{\delta}+\sqrt{\bar{\rho}})^2}{1-2\nu^2\bar{\rho}}\ge\frac{\nu^2\bar{\rho}}{4}$ for $\nu$ sufficiently small (e.g., $\nu<\frac{1}{8(\sqrt{\delta}+1)}$ suffices), and so $b\le\frac{4\delta}{\nu^2\bar{\rho}}m$ w.p. $\ge 1-e^{-Cm(1+o(1))}$.

[23]Write $\sigma_{min}^2(M)\;\ge\;\min_{\|\boldsymbol{x}_1\|_2^2+\|\boldsymbol{x}_2\|_2^2=1}\left(\|A\boldsymbol{x}_1\|_2-\|B\boldsymbol{x}_2\|_2\right)^2+\|\boldsymbol{x}_2\|_2^2$. Setting $\lambda=\|\boldsymbol{x}_1\|_2^2$, the previous is $\ge$ $\min_{\lambda\in[0,1]}\ \sigma_{min}^2(A)+(1-\sigma_{min}^2(A))(1-\lambda)-2\|A\|\|B\|\sqrt{1-\lambda}$, which is minimized at $\sqrt{1-\lambda}=\frac{\|A\|\|B\|}{1-\sigma_{min}^2(A)}$.

For $c = \mathbf{1}^* Q^{-1}\boldsymbol{\zeta}$, notice that $\boldsymbol{\zeta} = Z_{J^c,\bullet}^* \boldsymbol{\mu}_{J^c}$ is iid $\mathcal{N}(0, \nu^2\alpha/m)$. Write $Q = Z_{J^c,\bullet}^* \pi_{\boldsymbol{\mu}_{J^c}^\perp} Z_{J^c,\bullet} + \left[\begin{smallmatrix} 0 & 0 \\ 0 & \mathtt{I} \end{smallmatrix}\right] + \frac{1}{\alpha}\boldsymbol{\zeta}\boldsymbol{\zeta}^* \doteq L + \frac{1}{\alpha}\,\boldsymbol{\zeta}\,\boldsymbol{\zeta}^*$, then $Q^{-1} = L^{-1} - L^{-1}\boldsymbol{\zeta}\frac{1}{\alpha+\boldsymbol{\zeta}^* L^{-1}\boldsymbol{\zeta}}\boldsymbol{\zeta}^* L^{-1}$, and $|\mathbf{1}^* Q^{-1}\boldsymbol{\zeta}| = \left|\mathbf{1}^* L^{-1}\boldsymbol{\zeta}\left(\frac{\alpha}{\alpha+\boldsymbol{\zeta}^* L^{-1}\boldsymbol{\zeta}}\right)\right| \leq |\mathbf{1}^* L^{-1}\boldsymbol{\zeta}|$. An identical argument[24] to the one given above for $Q$ shows that on the complement of an event of probability $\asymp e^{-Cm}$, $\sigma_{min}(L) \geq \frac{\nu^2\bar{\rho}}{4}$, and so $\|L^{-1}\mathbf{1}\|_2 \leq \frac{4\sqrt{\delta}}{\nu^2\bar{\rho}}\, m^{1/2}$. Since $\boldsymbol{\zeta}$ is independent of $L$, $\left\langle \frac{L^{-1}\mathbf{1}}{\|L^{-1}\mathbf{1}\|_2}, \boldsymbol{\zeta}\right\rangle$ is simply an $\mathcal{N}(0, \nu^2\alpha/m)$ random variable, and so for any $\varepsilon > 0$

$$P\left[\left|\mathbf{1}^* L^{-1}\boldsymbol{\zeta}\right| > \varepsilon m^{1/2}\right] \;\leq\; P\left[\|L^{-1}\mathbf{1}\| > \frac{4\sqrt{\delta}}{\nu^2\bar{\rho}} m^{1/2}\right] + P\left[\left|\left\langle \frac{L^{-1}\mathbf{1}}{\|L^{-1}\mathbf{1}\|_2}, \boldsymbol{\zeta}\right\rangle\right| > \varepsilon\frac{\nu^2\bar{\rho}}{4\sqrt{\delta}}\right] \;\asymp\; e^{-C_\varepsilon m}$$

for some constant $C_\varepsilon$ (where we have controlled the second part via standard Gaussian tail bounds[25]). So, with overwhelming probability, $|c| = |\mathbf{1}^* Q^{-1}\boldsymbol{\zeta}| \leq \varepsilon\, m^{1/2}$.

The final quadratic term is $d = \boldsymbol{\zeta}^* Q^{-1}\boldsymbol{\zeta} = \boldsymbol{\zeta}^* L^{-1}\boldsymbol{\zeta}\frac{\boldsymbol{\zeta}^* L^{-1}\boldsymbol{\zeta}}{\alpha+\boldsymbol{\zeta}^* L^{-1}\boldsymbol{\zeta}} \leq \boldsymbol{\zeta}^* L^{-1}\boldsymbol{\zeta}$. The norm of the $\delta m$-dimensional $\mathcal{N}(0, \nu^2\alpha/m)$ vector $\boldsymbol{\zeta}$ concentrates: by (21), $\|\boldsymbol{\zeta}\|_2 \leq \sqrt{2}\nu\sqrt{\alpha\delta}$ with probability at least $1 - e^{-Cm(1+o(1))}$. We exploit the fact that although $\|L^{-1}\| = O(\nu^{-2})$, for most vectors $L$ is well-conditioned (due to the presence of the identity matrix in $\left[\begin{smallmatrix} Z_1 & Z_2 \\ 0 & \mathtt{I} \end{smallmatrix}\right]$). Consider the subspace $\Sigma = \{\boldsymbol{x} \mid \boldsymbol{x}_I = 0\} \subset \mathbb{R}^n$. Since for all $\boldsymbol{x} \in \Sigma$, $\|L\boldsymbol{x}\|_2 \geq \|\boldsymbol{x}\|_2$, $\left\|L^{-1}\big|_{L\Sigma}\right\| \leq 1$, and

$$\begin{aligned}
\boldsymbol{\zeta}^* L^{-1}\boldsymbol{\zeta} &= \boldsymbol{\zeta}^*\left(L^{-1}\big|_{L\Sigma}\pi_{L\Sigma}\,\boldsymbol{\zeta} + L^{-1}\pi_{(L\Sigma)^\perp}\boldsymbol{\zeta}\right) \\
&\leq \|\boldsymbol{\zeta}\|_2^2\left\|L^{-1}\big|_{L\Sigma}\right\|_2 + \|L^{-1}\|_2\|\boldsymbol{\zeta}\|_2\left\|\pi_{(L\Sigma)^\perp}\boldsymbol{\zeta}\right\|_2 \quad\leq\quad 2\nu^2\alpha\delta + \frac{4\sqrt{2\alpha\delta}}{\nu\bar{\rho}}\left\|\pi_{(L\Sigma)^\perp}\boldsymbol{\zeta}\right\|_2.
\end{aligned}$$

The norm $\|\pi_{(L\Sigma)^\perp}\boldsymbol{\zeta}\|$ of the projection of $\boldsymbol{\zeta}$ onto an independent $k_1$-dimensional subspace is distributed as the norm of a $k_1$-dimensional $\mathcal{N}(0, \nu^2\alpha/m)$ vector: $P\left[\left\|\pi_{(L\Sigma)^\perp}\boldsymbol{y}\right\| \geq \varepsilon'\nu\sqrt{\alpha}\right] \asymp e^{-2\varepsilon'^2 m/\pi^2}$. For appropriate $\varepsilon$, with overwhelming probability, $d \leq \boldsymbol{\zeta}^* L^{-1}\boldsymbol{\zeta} \leq 4\nu^2\alpha\delta$.

The denominator of $\Xi$ in (52) is $b(\alpha-d)+(c+1)^2 \geq C_b\alpha(1-4\nu^2\delta)m$. By Lemma 6, $\alpha = \|\boldsymbol{\mu}_{J^c}\|_2^2 \geq \bar{\rho}/2$ w.p. $\geq 1 - e^{-Cm(1+o(1))}$, and so the denominator is $\geq C_{denom}m$ with overwhelming probability. Since each of the terms in the numerator is $\leq Cm$ with overwhelming probability, $\|\Xi\| \leq C_\Xi$ for appropriate constant $C_\Xi$. Since the columns of $M$ have unit norm, $\|M\| \leq 2$, and

$$\|(G^*G)^{-1}\| \;\leq\; \|Q^{-1}\| + \|Q^{-1}\|\,\|M\|^2\,\|\Xi\| \;\leq\; \frac{4}{\nu^2\bar{\rho}} + \frac{4}{\nu^2\bar{\rho}}\,4\,C_\Xi \;\doteq\; C_G,$$

a constant, establishing the first assertion of the lemma.

[24]Consider instead $\sigma_{min}^2\left(\begin{bmatrix} \pi_{\boldsymbol{\mu}_{J^c}^\perp} Z_1 & \pi_{\boldsymbol{\mu}_{J^c}^\perp} Z_2 \\ 0 & \mathtt{I} \end{bmatrix}\right)$. The singular values of $\pi_{\boldsymbol{\mu}_{J^c}^\perp} Z_2$ are distributed as those of a $(\bar{\rho}m - 1) \times (\delta m - k_1)$ iid $\mathcal{N}(0, \nu^2/m)$ matrix. The bounds given by Fact 1 are essentially the same as those for $Z_2$.

[25]For example, if $X$ is $\mathcal{N}(0, \sigma^2)$, $P\left[|X| \geq \sigma t\right] \leq t^{-1}e^{-t^2/2}$.

We next extend the above reasoning to bound $(G^*G)^{-1}\mathbf{1}$ and $\mathbf{1}^*(G^*G)^{-1}\mathbf{1}$. Notice that

$$(G^*G)^{-1}\mathbf{1} \;=\; \frac{c+1}{b\,(\alpha-d)+(c+1)^2}\,Q^{-1}\mathbf{1} \;-\; \frac{1}{\alpha-d+(c+1)^2/b}\,Q^{-1}\boldsymbol{\zeta} \;\doteq\; \lambda_1\,Q^{-1}\mathbf{1}+\lambda_2\,Q^{-1}\boldsymbol{\zeta}.$$

For any $\varepsilon>0$, $|\lambda_1| \;\le\; \frac{|\,c+1\,|}{b\,(\alpha-d)} \;\le\; \frac{\varepsilon\,m^{1/2}+1}{C_b\,m\,\frac{\rho}{2}\,(1-4\nu^2\delta)}$ with overwhelming probability. Hence for any $\varepsilon''>0$, $|\lambda_1| \le \varepsilon'' m^{-1/2}$ for $m$ sufficiently large, on the complement of a bad event of probability $\asymp e^{-Cm}$. Similarly, $|\lambda_2| \le \frac{1}{\alpha-d} \le \frac{2}{\bar\rho\,(1-4\nu^2\delta)}$, and so

$$\|(G^*G)^{-1}\mathbf{1}\|_2 \;\le\; |\lambda_1|\|Q^{-1}\|\|\mathbf{1}\| + |\lambda_2|\|Q^{-1}\|\|\boldsymbol{\zeta}\| \;\le\; \frac{4\,\varepsilon''\sqrt{\delta}}{\nu^2\bar\rho} + \frac{8\sqrt{2\delta}}{\nu^2\bar\rho^2\,(1-4\nu^2\delta)} \;\doteq\; C_1.$$

Similarly, $\mathbf{1}^*(G^*G)^{-1}\mathbf{1} \;=\; \frac{b}{b(\alpha-d)+(c+1)^2} \;\le\; \frac{2}{\bar\rho\,(1-4\nu^2\delta)} \;\doteq\; C_2$.

We need one more bound, for $|\langle \boldsymbol{\mu}_J, \boldsymbol{\sigma}\rangle|$. Consider the Martingale $(X_i)_{i=0}^{\rho m}$ given by $X_0=0$, $X_i = \sum_{j=1}^{i} \boldsymbol{\mu}_J(j)\boldsymbol{\sigma}(j)$. We are interested in $X_{\rho m} = \langle \boldsymbol{\mu}_J, \boldsymbol{\sigma}\rangle$. Since $|X_i - X_{i-1}| \le |\boldsymbol{\mu}_J(i)|$, by Hoeffding's inequality [31],

$$P\left[\,|X_{\rho m}| \ge t\,\right] \;\le\; 2\exp\left(-\frac{t^2}{2\sum_{j=1}^{\rho m}\boldsymbol{\mu}_{J(j)}^2}\right) \;\le\; 2\,e^{-\frac{t^2}{2}}, \tag{54}$$

and so with probability $\ge 1-e^{-Cm^{1-\eta_0/2}}$, $|\langle\boldsymbol{\mu}_J,\boldsymbol{\sigma}\rangle| \le m^{1/2-\eta_0/4}$.

With these results in hand, recall that

$$\begin{aligned}\boldsymbol{q}_0 \;=\;& \left[\begin{smallmatrix} Z_1 & Z_2 \\ 0 & \mathrm{I}\end{smallmatrix}\right](G^*G)^{-1}Z_{J,\bullet}^*\boldsymbol{\sigma} \;+\; \left[\begin{smallmatrix} Z_1 & Z_2 \\ 0 & \mathrm{I}\end{smallmatrix}\right]\Big(-(G^*G)^{-1}\mathbf{1}_I + \langle\boldsymbol{\mu}_J,\boldsymbol{\sigma}\rangle\,(G^*G)^{-1}\mathbf{1}\Big) \\ &+\; \left[\begin{smallmatrix} \boldsymbol{\mu}_{J^c} \\ \mathbf{0}\end{smallmatrix}\right]\Big(-\mathbf{1}^*(G^*G)^{-1}\mathbf{1}_I \;+\; \langle\boldsymbol{\mu}_J,\boldsymbol{\sigma}\rangle\,\mathbf{1}^*(G^*G)^{-1}\mathbf{1}\Big) + \left[\begin{smallmatrix} \boldsymbol{\mu}_{J^c} \\ \mathbf{0}\end{smallmatrix}\right]\mathbf{1}^*(G^*G)^{-1}Z_{J,\bullet}^*\boldsymbol{\sigma}.\end{aligned} \tag{55}$$

The second term of (55), $\left\|\left[\begin{smallmatrix} Z_1 & Z_2 \\ 0 & \mathrm{I}\end{smallmatrix}\right]\Big(-(G^*G)^{-1}\mathbf{1}_I + \langle\boldsymbol{\mu}_J,\boldsymbol{\sigma}\rangle\,(G^*G)^{-1}\mathbf{1}\Big)\right\|$ is bounded above by

$$\Big(1+\sqrt{2}\nu(\sqrt{\delta}+\sqrt{\bar\rho})\Big)\Big(C_G\sqrt{C_0}\,m^{1/2-\eta_0/2} + m^{1/2-\eta_0/4}C_1\Big)$$

w.p. $\ge 1-e^{-Cm^{1-\eta_0/2}(1+o(1))}$. Similarly, for the third term of (55)

$$\left\|\left[\begin{smallmatrix} \boldsymbol{\mu}_{J^c} \\ \mathbf{0}\end{smallmatrix}\right]\Big(-\mathbf{1}^*(G^*G)^{-1}\mathbf{1}_I \;+\; \langle\boldsymbol{\mu}_J,\boldsymbol{\sigma}\rangle\,\mathbf{1}^*(G^*G)^{-1}\mathbf{1}\Big)\right\| \;\le\; C_1C_0m^{1/2-\eta_0/4} + C_2\,m^{1/2-\eta_0/4}.$$

For the final term of (55), $\boldsymbol{\vartheta} \doteq Z_{J,\bullet}^*\boldsymbol{\sigma}$ is distributed as an iid $\mathcal{N}(0,\nu^2\rho)$ vector, independent of $G$, and so

$$P\left[\left|\left\langle \frac{(G^*G)^{-1}\mathbf{1}}{\|(G^*G)^{-1}\mathbf{1}\|}, \boldsymbol{\vartheta}\right\rangle\right| \ge m^{1/2-\eta_0/4}\right] \asymp e^{-Cm^{1-\eta_0/2}}. \tag{56}$$

On the complement of this bad event,

$$\left\|\boldsymbol{\mu}_{J^c}\mathbf{1}^*(G^*G)^{-1}\boldsymbol{\vartheta}\right\| \;\le\; \|(G^*G)^{-1}\mathbf{1}\|\cdot\left|\left\langle \frac{(G^*G)^{-1}\mathbf{1}}{\|(G^*G)^{-1}\mathbf{1}\|}, \boldsymbol{\vartheta}\right\rangle\right| \;\le\; C_1\,m^{1/2-\eta_0/4}. \tag{57}$$

■

*Lemma 8 (Concentration for Gaussian tops):* Fix $\sigma \leq 1$, $\varepsilon \leq 1/2$. Let $\boldsymbol{x}$ be a $d$-dimensional random vector with entries iid $\mathcal{N}(0, \sigma^2)$, and let $\theta$ be the operator that takes the part of $\boldsymbol{x}$ above $1-\varepsilon$:

$$\theta : \mathbb{R}^d \to \mathbb{R}^d \text{ such that } [\theta \boldsymbol{x}](i) = \begin{cases} \operatorname{sgn}(\boldsymbol{x}(i))(|\boldsymbol{x}(i)| - 1 + \varepsilon), & |\boldsymbol{x}(i)| > 1 - \varepsilon, \\ 0, & \text{else.} \end{cases} \tag{58}$$

Then $P\left[\, \|\theta \boldsymbol{x}\|_2 \geq 4e^{-\frac{1}{16\sigma^2}}\, d^{1/2} \right] \asymp e^{-C_\sigma d}$, where $C_\sigma$ is a constant (w.r.t. $d$) depending only on $\sigma$.

*Proof:* Let $\boldsymbol{y} \in \mathbb{R}^d$ be iid $\mathcal{N}(0,1)$, then $\|\theta\boldsymbol{x}\|_2$ is equal in distribution to $\|\theta\sigma\boldsymbol{y}\|_2$. Now, $\mathbb{E}\|\theta\sigma\boldsymbol{y}\|_2^2 = d \cdot \mathbb{E}(\theta\boldsymbol{x}(i))^2 = \frac{d}{\sigma}\sqrt{\frac{2}{\pi}} \int_{1-\varepsilon}^{\infty} t^2 e^{-t^2/2\sigma^2} dt$. Integrating by parts[26] yields

$$d^{-1}\mathbb{E}\|\theta\sigma\boldsymbol{y}\|_2^2 \;=\; \frac{(1-\varepsilon)\,\sigma}{\sqrt{\pi/2}} e^{-\frac{(1-\varepsilon)^2}{2\sigma^2}} + 2\,\sigma^2 Q\left(\tfrac{1-\varepsilon}{\sigma}\right) \;\leq\; \sigma\sqrt{\frac{2}{\pi}}\,\frac{1+\sigma^2}{1-\varepsilon}\, e^{-\frac{(1-\varepsilon)^2}{2\sigma^2}} \;\leq\; 4\sigma e^{-\frac{1}{8\sigma^2}},$$

and $\mathbb{E}[\,\|\theta\sigma\boldsymbol{y}\|_2\,] \leq 2\,e^{-\frac{1}{16\sigma^2}} d^{1/2}$. Meanwhile, $\mathbb{E}\sqrt{\sum_{i=1}^d |\theta\sigma\boldsymbol{y}(i)|^2} = \sqrt{d}\;\mathbb{E}\sqrt{\frac{\sum_{i=1}^d |\theta\sigma\boldsymbol{y}(i)|^2}{d}}$. It is not difficult to show[27] that $\mathbb{E}\sqrt{\frac{\sum_{i=1}^d |\theta\sigma\boldsymbol{y}(i)|^2}{d}} \to C'_\sigma$ for some constant $C'_\sigma > 0$, and so $\mathbb{E}\|\theta\sigma\boldsymbol{y}\|_2 \geq C'_\sigma\, d^{1/2}$. Since $f(\cdot) = \|\theta\sigma \cdot\|_2$ is 1-Lipschitz for $\sigma \leq 1$, $P\left[\,\|\theta\sigma\boldsymbol{y}\|_2 \geq 2\,\mathbb{E}\|\theta\sigma\boldsymbol{y}\|_2\,\right] \leq \exp\left(-8(\mathbb{E}\|\theta\sigma\boldsymbol{y}\|_2)^2/\pi^2\right)$ [3]. Plugging in the upper and lower bounds on $\mathbb{E}|\theta\sigma\boldsymbol{y}\|_2$ yields the result. ■

### *C. Details of the Proof of Theorem 1*

*Proof:* Consider the weak proportional growth setting $\mathrm{WPG}_{\delta,\rho,C_0,\eta_0}$ with $\bar{\rho} < \delta$. We first consider a fixed, arbitrary sequence of signal supports $I \in \binom{[n]}{k_1}$. By Lemma 2, $(I, J, \boldsymbol{\sigma})$ is $\ell^1$-recoverable if $\exists\, c \in (0,1)$ such that

$$\|\boldsymbol{q}_0\|_2 + \frac{1}{1-\xi}\|\theta\boldsymbol{q}_0\|_2 \;\leq\; (1-\varepsilon)\sqrt{cp} \;=\; (1-\varepsilon)\sqrt{c\,(\bar{\rho}+\delta)}\, m^{1/2} + o(m^{1/2}), \tag{59}$$

where $\xi = \inf_{\|\boldsymbol{s}\|_0 \leq cp} \|\pi_{\mathcal{R}(G)}\boldsymbol{s}\|_2/\|\boldsymbol{s}\|_2$. Choose $c$ small enough that $\beta \doteq (\bar{\rho}+\delta)c$ satisfies $\beta < \min\left(\frac{\bar{\rho}}{1024}, \frac{\bar{\rho}}{64\,(1+2C_\mu\bar{\rho}^{-1/2})^2}\right)$ and $\bar{\rho}H(\beta/\bar{\rho}) + \delta H(\beta/\delta) < \frac{\bar{\rho}}{128\pi^2}$ (since in Lemma 3, $\|\boldsymbol{s}\|_0$ is a fraction of $m$, not $p$). Further suppose that $\nu < \min(\frac{1}{9}, \frac{1}{8(\sqrt{\delta}+1)}, (512/\delta)^{-1/4})$. Then by Lemma 3, $\xi < 1 - C_\xi \nu^8$, with probability $1 - e^{-Cm(1+o(1))}$.

Meanwhile, by Lemma 4, with probability at least $1 - e^{-Cm^{1-\eta_0/2}(1+o(1))}$, $\|\boldsymbol{q}_0\|_2 \leq \alpha_1\nu m^{1/2} + o(m^{1/2})$ and $\|\theta\boldsymbol{q}_0\| \leq \alpha_2\nu^{-8}e^{-\frac{1}{64\nu^2}}$. On the intersection of these three good events, the left hand side of (59) becomes

$$\|\boldsymbol{q}_0\|_2 + \frac{1}{1-\xi}\|\theta\boldsymbol{q}_0\| \;\leq\; \alpha_1\,\nu\, m^{1/2} + \alpha_2\nu^{-8}\exp\left(-\frac{1}{64\nu^2}\right) m^{1/2} + o(m^{1/2}). \tag{60}$$

[26]And noting that $Q(z) \leq \frac{1}{z\sqrt{2\pi}} e^{-z^2/2}$.

[27]Apply the strong law of large numbers to $d^{-1}\sum|\theta\sigma\boldsymbol{y}(i)|^2$ and Slutsky's theorem (Theorem 6 of [32]) to argue that $\mathbb{E}\sqrt{d^{-1}\sum|\theta\sigma\boldsymbol{y}(i)|^2} \to \sqrt{\mathbb{E}|\theta\sigma\boldsymbol{y}(i)|^2}$.

For $\nu$ sufficiently small, this is $\leq (1-\varepsilon)\sqrt{c(\bar{\rho}+\delta)}m^{1/2} + o(m^{1/2})$, and hence, for $m$ sufficiently large, on an event of probability $\geq 1 - \exp\left(-Cm^{1-\eta_0/2}(1+o(1))\right)$, $(I, J, \boldsymbol{\sigma})$ is $\ell^1$-recoverable. There are $\binom{m}{k_1} \leq \exp(m^{1-\eta_0}\log m)$ subsets $I$, and so the probability that $(I, J, \boldsymbol{\sigma})$ is not $\ell^1$-recoverable for some $I$ is bounded by

$$\exp\left(-Cm^{1-\eta_0/2}(1+o(1))\right) \times \exp\left(m^{1-\eta_0}\log m\right) \quad = \quad \exp\left(-Cm^{1-\eta_0/2}(1+o(1))\right) \quad = \quad o(1),$$

establishing the theorem. ■